\documentclass[aps, prd, twocolumn, showpacs, a4paper,
nofootinbib, superscriptaddress,longbibliography, floatfix]{revtex4-2}
\usepackage{blindtext}
\usepackage{graphicx}
\usepackage{epsfig}
\usepackage{bm}
\usepackage{amsfonts}
\usepackage{pgf}
\usepackage{color}
\usepackage[T1]{fontenc}
\usepackage[latin1]{inputenc}
\usepackage{graphicx}
\usepackage[english]{babel}
\usepackage{amsmath}
\usepackage{amssymb}
\usepackage{amsfonts}
\usepackage{makecell}
\usepackage{hyperref}
\usepackage[draft,deletedmarkup=xout]{changes}
\usepackage{mathtools}
\usepackage{xcolor}
\usepackage{multirow}
\usepackage[version=4]{mhchem}
\usepackage{soul}
\definecolor{green}{rgb}{0.19,0.64,0.54}
\definecolor{blue}{rgb}{0,0,1}
\definecolor{reddish}{rgb}{0.65, 0.2, 0.2}
\definecolor{darkgreen}{rgb}{0.2,0.7,0.3}
\definecolor{darkblue}{rgb}{0.3,0.40,0.48}
\definecolor{gray}{rgb}{.8,.8,.8}

\hypersetup{,
	pdftitle={VOSCurr},
  	pdfauthor={I. Rybak et al.},
    colorlinks=true,       
    linkcolor=darkblue,          
    citecolor=darkgreen,        
    filecolor=reddish,      
    urlcolor=blue,           
    linktocpage=true
}
\usepackage{wrapfig}
\usepackage{lipsum}
\usepackage{ulem}

\graphicspath{ {./Figures/} }

\newcommand{\dd}{\mathrm{d}}

\newcommand{\cl}{c_\textsc{l}}
\newcommand{\ct}{c_\textsc{t}}

\newcommand{\Rez}{\Re\mathrm{e}}

\newcommand{\sca}{_\textsc{sc}}

\begin{document}

\title{Charge-velocity-dependent one-scale linear model}

\author{C. J. A. P. Martins}
\email{Carlos.Martins@astro.up.pt}
\affiliation{Centro de Astrof\'{\i}sica da Universidade do Porto, Rua das
Estrelas, 4150-762 Porto, Portugal}
\affiliation{Instituto de Astrof\'{\i}sica e Ci\^encias do Espa\c co,
CAUP, Rua das Estrelas, 4150-762 Porto, Portugal}

\author{Patrick Peter}
\email{peter@iap.fr}
\affiliation{${\cal G}\mathbb{R}\varepsilon\mathbb{C}{\cal O}$ -- Institut
d'Astrophysique de Paris, CNRS \& Sorbonne Universit\'e, UMR 7095
98 bis boulevard Arago, 75014 Paris, France}
\affiliation{Centre for Theoretical Cosmology, Department of Applied
Mathematics and Theoretical Physics, University of Cambridge, Wilberforce
Road, Cambridge CB3 0WA, United Kingdom}

\author{I. Yu. Rybak}
\email[]{Ivan.Rybak@astro.up.pt}
\affiliation{Centro de Astrof\'{\i}sica da Universidade do Porto, Rua das
Estrelas, 4150-762 Porto, Portugal}
\affiliation{Instituto de Astrof\'{\i}sica e Ci\^encias do Espa\c co,
CAUP, Rua das Estrelas, 4150-762 Porto, Portugal}

\author{E. P. S. Shellard}
\email{E.P.S.Shellard@damtp.cam.ac.uk}
\affiliation{Centre for Theoretical Cosmology, Department of Applied
Mathematics and Theoretical Physics, University of Cambridge, Wilberforce
Road, Cambridge CB3 0WA, United Kingdom}

\begin{abstract}
We apply a recently developed formalism to study the evolution of a
current-carrying string network under the simple but generic
assumption of a linear equation of state.  We demonstrate that the
existence of a scaling solution with non-trivial current depends on
the expansion rate of the universe, the initial root mean square
current on the string, and the available energy loss mechanisms. We
find that the fast expansion rate after radiation-matter equality will
tend to rapidly dilute any pre-existing current and the network will
evolve towards the standard Nambu-Goto scaling solution (provided
there are no external current-generating mechanisms). During the
radiation era, current growth is possible provided the initial
conditions for the network generate a relatively large current and/or
there is significant early string damping.  The network can then
achieve scaling with a stable non-trivial current, assuming large
currents will be regulated by some leakage mechanism. The potential
existence of current-carrying string networks in the radiation era,
unlike the standard Nambu-Goto networks expected in the matter era,
could have interesting phenomenological consequences.
\end{abstract}

\date{\today}
\maketitle

\section{Introduction}

Models of the early universe suggest that the symmetry breaking phase
transitions can lead to the formation of one-dimensional topological
defects, known as cosmic strings \cite{HindmarshKibble,
  Vilenkin:2000jqa, CopelandKibble}.

An evolving network of these objects will leave observationally
discernible features, including anisotropies in the cosmic microwave
background \cite{LazanuShellardLandriau, LazanauShellard,
  CharnockAvgoustidisCopelandMoss,
  LizarragaUrrestillaDaverioHindmarshKunz,
  HindmarshLizarragaUrrestillaDaverioKunz, RybakSousa}, gravitational
lensing \cite{Sazhin1, Sazhin2} and a stochastic background of
gravitational waves
\cite{Blanco-PilladoOlum,LISA,SousaAvelinoGuede,LIGO}, which can be
probed by present and future observational programmes. However,
accurate observational predictions of such cosmic string features can
only be obtained once one has a quantitatively accurate understanding
of their evolution.  Several analytic models, with different levels of
detail, are available to describe the evolution of the string networks
\cite{KIBBLEEvolut, Bennett, AlbrechtTurok, AustinCopelandKibble,
  Martins:1996jp, Martins:2000cs, Vanchurin}.

The physical properties of cosmic strings are determined by the
specific details of the symmetry-breaking phase transition that
produced them. The Nambu-Goto (NG) and Abelian-Higgs models provide
the simplest descriptions, but one expects that physical realistic
cosmic strings have additional degrees of freedom. Of particular
interest are superconducting cosmic strings, which initially were
suggested in Ref.~\cite{Witten:1984eb}, since they are the expected
outcome of various high energy scenarios \cite{Peter:1993tm,
  DavisDavisTrodden, DavisPerkins, KibbleLozanoYates, GaraudVolkov};
it was even argued that when the low-energy limit of the
  string-forming model contains the standard
  SU(3)$\times$SU(2)$\times$U(1) model, currents must be produced, at
  least at the electroweak scale, but also at any intermediate scale
  structured in a similar way \cite{Davis:1995kk}.

The properties of superconducting strings have been previously studied
in Refs.~\cite{Peter:1992dw, Peter:1992ta, HartmannMichelPeter,
  HartmannMichelPeter2}.  These works suggest that effective models
with an analogous NG limit can approximately mimic superconducting
string behaviour \cite{Carter:1994hn, Carter:1999hx, Carter:2000wv,
  HartmannCarter}.  Nevertheless, a more systematic approach to the
evolution of these networks is not yet fully developed. Indeed,
  current-carrying degrees of freedom will substantially complicate
  the numerical study of cosmic string networks, which are already
  challenging for Nambu-Goto strings. Moreover, it would
  require entirely different algorithms, which would need to be
  developed \textit{ab initio}, rather than simply adapting an
  existing Nambu-Goto simulation, a problem which is compounded by
  increased memory requirements the potential need to accurately
  resolve new phenomena on very different lengthscales. In other
  words, the motivations for developing a reliable analytic model for
  the evolution of current and charge carrying string networks are
  even more compelling than those for the Nambu-Goto case.

Our approach to modeling the evolution of superconducting string
networks, is based on the canonical velocity-dependent one-scale (VOS)
model \cite{Martins:1996jp,Martins2016B}.  This approach already
demonstrated flexibility and usefulness in modeling particular
realizations of superconducting string networks \cite{Martins1998,
  MartinsShellard1998, Oliveira:2012nj, Martins:2014, Vieira:2016,
  Rybak:2017yfu}.  Here, we draw on the recently developed general VOS
formalism for superconducting string networks, which shall be referred
to as charged or current-carrying VOS (CVOS), developed in
Ref.~\cite{MPRS}, and explore some of its consequences. Specifically,
the aim of this sequel is to study in detail the dynamics of
superconducting cosmic string networks with a linear equation of state
and to determine conditions when a scaling solution with non-trivial
charge amplitude is possible.

We emphasize that while the assumption of a linear equation of state
may seem too simple, it is not merely used for the sake of
computational simplicity. Instead, it should be a good approximation
when one has small charges and currents on the strings, which is
expected to be generically the case for string-forming phase
transitions in the early universe \cite{Witten:1984eb}. On the
  other hand, the need to accurately describe the additional degrees
  of freedom evolving on the string worldsheet implies that the CVOS
  model will be more complex than its simpler VOS counterpart. The
  larger number of coupled evolution equations yields richer dynamical
  phenomena.  This is unavoidable if one wants an accurate,
  self-consistent and quantitative description, but we also emphasize
  that this averaged (macroscopic) approach is nevertheless far
  simpler than any analogous microscopic study, let alone direct
  numerical simulations. Admittedly for the simpler string models,
  numerical simulations play an important role in calibrating (or at
  least providing bounds on) the analytic phenomenological parameters,
  which is a source of information which is not yet available for the
  CVOS model. However, this motivates our present approach of
  systematically studying the possible scaling behaviours of these
  networks (and in particular the behaviour of the charges and
  currents) under various possible assumptions for the model's
  phenomenological parameters, so that we can at least develop a
  robust qualitative understanding.

Our analysis herein may be thought of as a stress test of the model's
scaling solutions, in the sense that we will assume that asymptotic
scaling solutions with non-zero charge exist and study their form
under various assumptions for the available energy loss
mechanisms. Broadly speaking, we find that such solutions are
physically problematic, except in limited regions of parameter
space. In other words, these generalized scaling solutions are
possible but not generic. Indeed, two other types of solution also
exist, and are more common. The first of these recovers the standard
NG scaling solution, while the second are solutions where the charge
and current grow on the strings and would eventually lead to a
`frozen' network. The generalized scaling solutions can therefore be
thought of as equilibrium points between charge growth and its
disappearance. Several factors impact the behaviour of the network,
with the most crucial one being the expansion rate of the universe:
the slower the expansion rate, the more likely it is that charges and
currents survive on the strings. We study generic universes where the
scale factor grows as a power of physical or conformal time and find,
broadly speaking and for the physically expected values of the model
parameters, that fast expansion rates (including the matter dominated
epoch) lead to the Nambu-Goto solution while slow expansion rates
(including the radiation dominated epoch) can lead to charge growth or
scaling.  These three types of solution have previously been
identified for chiral superconducting strings \cite{Oliveira:2012nj},
and more recently also for wiggly strings \cite{Almeida}.  Finally,
through a stability analysis, we also identify another key factor--the
initial conditions--that determine which of these dynamical outcomes
for the network is physically realised. A scaling solution with
non-trivial charge (or a growing charge solution) appears only to be
possible for an initial state in which the current or charge density
is significant (and/or the velocities are low).

\section{Macroscopic model}
\label{I}

In previous work~\cite{MPRS}, we proposed a formalism extending the
VOS model to include current-carrying string networks (often referred
to as superconducting) which exhibit microscopic dynamics on the
string worldsheet described by a given equation of state (that is, the
energy per unit length and tension specified as functions of a state
parameter). This approach describes the network properties in terms of
averaged quantities, in particular the root mean square (RMS) velocity
$v$ and the correlation length in comoving units $\xi_\text{c}$ are
sufficient to encode the Nambu-Goto network properties. Here, however,
these must be supplemented with the averaged (timelike) charge $Q$ and
(spacelike) current $J$ flowing along the strings, which are
essentially the RMS averages of the corresponding timelike and
spacelike components of the microscopic current
respectively~\cite{MPRS}. It proves convenient to use the averaged
Lorentz-invariant two-current amplitude
\begin{equation}
\label{Chirality}
K = Q^2 - J^2,
\end{equation}
and to define the relative energy density due 
to the current and charge as 
\begin{equation}
\label{Charge}
Y = {\textstyle\frac{1}{2}}(Q^2 + J^2)\,.
\end{equation}
As the limit $K\to 0$ has been dubbed ``chiral'' in previous works
\cite{Carter:1999hx,Davis:2000cx}, we shall in what follows refer to
$K$ as the distance to chirality or, for the sake of conciseness, the
     {\sl chirality}. We shall describe $Y$ as the relative {\sl
       charge}, encompassing the effective energy density trapped in
     both the string current and charges. Adding this internal energy
     to that of the bare string $E_0$ one can define the total energy
     $E$ in a volume $V$ and thereby extend the definition of the
     string characteristic length, $L_\textsc{c}$, expressing in
     comoving units, as
\begin{equation}
\label{CurrCharLength}
E = \frac{\mu_0 V}{a^2 L_\textsc{c}^2},
\end{equation}
where we continue to make a Brownian assumption for network
correlations.  Note that in the standard VOS model the string
characteristic length, can be assumed to coincide with the string
correlation length, but that assumption is clearly not applicable
here---in other words, we will no longer have a one-scale model.

\subsection{Linear model}
\label{linearModel}

Our extended CVOS model is based on several assumptions, already to be
found in the structureless original VOS model; given that the model
quantitatively describes many of the characteristic features of
Nambu-Goto string networks, one can expect its current-carrying
extension to similarly reproduce superconducting cosmic string network
properties.  The assumptions are that the microscopic variables are
uncorrelated when averaged along the string, that they satisfy
specific boundary conditions encoding a torus-like topology for the
universe (on scales much larger than the observed one), and that the
string network, having been stochastically produced through a phase
transition, is therefore Brownian.  We will also consider additional
charge and energy loss mechanisms below, that is, beyond those due to
direct loop production.

The most important ingredient that explicitly appears in the equations
for our macroscopic variables is the underlying microscopic equation
of state, which was also assumed to translate into a macroscopic
one. Current-carrying strings are here assumed infinitely thin (after
integration of the transverse degrees of freedom), with a local state
parameter $\kappa$, made from the gradient of a scalar field (often a
phase in the underlying microscopic theory) living on the
2-dimensional worldsheet. The local energy per unit length and tension
of the string are calculated by means of a 2-dimensional surface
Lagrangian $f(\kappa)$. Once integrated (averaged) over the full
network, the state parameter $\kappa$ yields the time-dependent
chirality $K$, and the Lagrangian turns into a function of $K$, called
$F(K)$, encoding the microphysics.

In the present work, we explore the consequences of the extended VOS
model under the special assumption that the equation of state is of
the linear kind, as is expected to be appropriate in the limit of
small currents. In other words, we have
\begin{equation}
\label{FLinear}
F(K) = 1 - \frac{\kappa_0}{2}  K,
\end{equation}
where $\kappa_0$ is a positive constant. We will comment on the
applicability of our results to more generic equations of state in the
concluding section. The equation of state \eqref{FLinear} can be
understood as a linear correction to the NG action
\cite{Witten:1984eb, Carter:1999hx}, i.e., the lowest order expansion
of a general function $F(K)$, valid for small values of the current
$K\ll 1$. For this linear model \eqref{FLinear}, the averaged tension
$T$ and energy per unit length $U$ are given by \cite{Carter:1994hn,
  Rybak:2017yfu},
\begin{equation}
\begin{gathered}
\label{UTLinear}
U = \mu_0 \left( 1 + \frac{\kappa_0}{2} |K| \right), \\
T = \mu_0 \left( 1 - \frac{\kappa_0}{2} |K| \right),
\end{gathered}
\end{equation}
where $\mu_0$ is a constant with units of mass squared.

One could argue against the use of such a model by considering the
transverse and longitudinal averaged velocities of perturbations
propagation, respectively given by
\begin{equation}
\label{cLcT}
\ct^2 \equiv \frac{T}{U} =
\frac{ 1 - |\kappa_0 K| / 2 }{1 + |\kappa_0 K|/2}
\ \ \ \hbox{and}\ \ \ 
\cl^2 \equiv -\frac{\dd T}{\dd U} = 1,
\end{equation}
implying that this is a subsonic ($\cl>\ct$) type of string, while the
field theory of the original $U(1) \times U(1)$ Witten model,
currently accepted as standard to describe superconducting cosmic
strings, has been shown to be of the supersonic ($\ct>\cl$)
type~\cite{Peter:1992dw}. Other models of potential cosmological
relevance, for instance that obtained by integrating out the
short-scale wiggles to consider only smooth strings with a non-trivial
equation of state~\cite{Martin:1995xh,Martins1998}, have been shown to
be transonic~\cite{Carter:1994hn}.

The super- or subsonicity property however, is known to be relevant
for the stability~\cite{Martin:1994jp} of would-be
vortons~\cite{Davis:1988ij}. It may have important consequences on the
trajectories of individuals strings, such as enhancing loop formation,
energy losses or even charge
leakage~\cite{MartinPeter,CorderoCid:2002ts}.  As such effects are
already taken care of at the phenomenological level, with parameters
in principle to be determined by comparison with yet-to-be-done
numerical simulations, our conclusions should hold for these models
provided the equation of state itself remains, on average, in the
linear regime.

\subsection{Cosmological setup}

The string network evolution will be studied on a flat
Friedmann-Lema\^{\i}tre-Robertson-Walker (FLRW) background with line
element
\begin{equation}
\dd s_\textsc{flrw}^2 = a^2(\tau) \left(\dd\tau^2 - \dd \bm{x}^2\right),
\label{FLRW}
\end{equation}
where $\tau$ is the conformal time and $a$ the scale factor.  In
Ref.~\cite{MPRS}, the extended VOS model with an arbitrary equation of
state in such a FLRW background was derived; using the linear equation
of state \eqref{FLinear} reduces this system of equations to
\begin{subequations}
\label{EqOfMotMacroLin}
\begin{align}
\label{EqOfMotMacroLinA}
\dot{L}_\mathrm{c} = &\ \frac{\dot{a}}{a}\frac{L_\mathrm{c}}{1+ \kappa_0 Y}
\left(
v^2 + \kappa_0 Y \right) + \frac{g c v}{2\sqrt{1+ \kappa_0 Y}}, \\
\label{EqOfMotMacroLinB} \dot{v}  = &\ \frac{1-v^2}{1+\kappa_0 Y}
\left[ \frac{(1- \kappa_0 Y) k}{L_\mathrm{c}\sqrt{1+Y}} - 2 v
\frac{\dot{a}}{a} \right],  \\
\label{EqOfMotMacroLinC} \kappa_0 \dot Y = &\ 2 \kappa_0 Y
\left( \frac{v k}{L_\mathrm{c}
  \sqrt{1+ \kappa_0 Y}} - \frac{\dot{a}}{a} \right) \nonumber  \\
&  - \frac{v}{L_\mathrm{c}} c \left( g - 1
\right) \sqrt{1+ \kappa_0 Y}, \\
 \kappa_0 \dot K = &\ 2 K \kappa_0 \left( \frac{v k}{L_\textsc{c}
\sqrt{1+ \kappa_0 Y}} - \frac{\dot{a}}{a} \right) \nonumber \\
&- 2 \frac{v}{L_\mathrm{c}} c \left( g - 1
\right) \left(1-2\rho\right)\sqrt{1+ \kappa_0 Y}, 
\label{EqOfMotMacroLinD}
\end{align}
\end{subequations}
where a dot denotes differentiation with respect to $\tau$
($\dot{A}\equiv \dd A/\dd \tau$), the chirality is as defined above
\eqref{Chirality}, along with the relative charge \eqref{Charge},
while the relation between the characteristic lengths $\xi_\textsc{c}$
and $L_\textsc{c}$ for a linear equation of state \eqref{FLinear} is
given by
\begin{equation}
    \label{LXi}
    \xi_\textsc{c} = L_\textsc{c} \sqrt{1+\kappa_0 Y}.
\end{equation}
One can immediately see that by rescaling $K \rightarrow K/\kappa_0$
and $Y \rightarrow Y/\kappa_0$, one can absorb the $\kappa_0$
dependence in the whole system; hence, and without loss of generality,
we shall set the constant $\kappa_0\to1$ from now on, as this amounts
to redefining the units in which the charge and chirality are
measured.

There are six parameters we have introduced in the CVOS evolution
equations, with four of these in Eq.~\eqref{EqOfMotMacroLin}.  The
first two essentially govern the dynamics of the NG network in the
original VOS model \eqref{EqOfMotMacroLin}, namely:
\begin{itemize}
\item[(i)] the momentum parameter $k$ is, in principle, a function of
  the RMS velocity $k=k(v)$, representing the averaged scalar product
  of the string velocity and normalized curvature vectors (discussed
  at length in Ref.~\cite{MPRS}), and
\item[(ii)] the loop chopping efficiency $c$ describing the key
  network energy loss mechanism.
\end{itemize}
To these we add four new physically-motivated quantities determining
the loss mechanisms for current and charge during network evolution:
\begin{itemize}
\item[(iii)] the current chopping efficiency $g$ accounting for
  whether loops typically lose above or below the average current and
  charge from the network, together with the bias or relative
  proportion $\rho$ of the current versus the charge that escapes with
  each loop,
\item[(iv)] a charge and current loss parameter $A$ describing direct
  leakage from long strings and the corresponding bias
  $\rho_\textsc{a}$ between the relative current and charge that
  leaves the string.
\end{itemize}
This last parameter should in principle be evaluated from the
microscopic dynamics: when the curvature of the string leads to
direct leakage of the current, interactions with background
particles provide a model-dependent cross-section for ejecting
charged particles, or when string self-interactions such as
reconnection disrupt the string currents and cause further charge
losses \cite{IbeKobayashiNakayamaShirai}.  While we can
suggest the form of the dependence on macroscopic variables for
the charge leakage (see section \ref{IID}), we will not fix the
model-dependent parameter $A$, keeping our consideration generic.
For accurate estimation of the charge and current loss parameter
$A$, we would require a detailed analytic evaluation or numerical
simulations in the framework of a particular model. As no such
treatment is currently available, $A$ remains a purely
phenomenological and undetermined parameter.

We shall define these additional parameters in greater detail below,
but we note here that the generic network behaviours we find do not
appear to be particularly sensitive to the precise values of $g$,
$\rho$, $\rho_\textsc{a}$ and $A$.

\subsection{CVOS model parameters} \label{IB}

The extended, current-carrying, CVOS model includes phenomenological
terms describing energy losses of the long string network into loops.
Defining $E_0$ as the bare energy of the strings in a volume $V$ with
a correlation length (expressed in comoving units) $\xi_\textsc{c}$,
the energy stored in the network for a vanishing current contribution
is
\begin{equation}
\label{PhysCharLength}
E_0 = \frac{\mu_0 V}{a^2 \xi_\textsc{c}^2},
\end{equation}
thereby identifying the average (conformal) distance $\xi_\textsc{c}$
between strings.  Recall that in the VOS model, loop production is
modeled through the energy loss \cite{Martins:1996jp}
\begin{equation}
\label{BareEnergyLoss}
\frac{\dd E_0}{\dd \tau} \bigg|_{\text{loops}} = - c v \frac{E_0}{\xi_\textsc{c}},
\end{equation}
where the chopping efficiency $c$ encodes the typical energy lost by
the long string network into loops.

In addition to the bare string energy, however, the total energy $E$
of a string network must include the charge contribution, which can be
represented, for the linear equation of state, as \cite{MPRS}
\begin{equation}
\label{Energies}
E = E_0 \left( 1 + Y \right),
\end{equation}
This renormalisation results in the modified current-carrying
characteristic length $L_\textsc{c}$ defined in
\eqref{CurrCharLength}.  Following Eq.~\eqref{BareEnergyLoss}, the
complete energy loss for the total energy $E$ is modified but remains
in a similar form
\begin{equation}
\label{EnergyLoss}
\frac{\dd E}{\dd \tau} \bigg|_{\text{loops}} = - g c v \frac{E}{\xi_\textsc{c}},
\end{equation}
where we have introduced the current chopping efficiency $g$ to
correct the usual bare loop chopping efficiency $c$; given its
definition, $g$ is a parameter that represents how much of the charge
is lost by loops in comparison with infinite strings. In other words,
if $g<1$ (respectively $g>1$), there is typically less (resp. more)
charge on loops chopped off the network in comparison with the
remaining 'infinite' strings, the limiting case $g=1$ producing loops
containing the same amount of charge as the infinite strings.
Phenomenological analytic modeling of such charge biases has also been
considered \cite{Oliveira:2012nj}. Similarly to $c$, one naturally
expects that $g$ could depend on $Y$.

The charge $Y$ is directly affected by the energy loss term, given by
Eq.~(\ref{EnergyLoss}), since it is a sum of timelike and spacelike
components of the current. The chirality $K$, being defined as the
difference between the squared timelike and spacelike components, can
decrease or increase depending on which component of the current is
dominantly lost in the form of loops. In order to allow for this
possible bias between the losses in timelike and spacelike current
components, a parameter $0\leq \rho \leq 1$ is introduced: the
unbiased case is represented by the midpoint $\rho_\text{\st{bias}} =
1/2$. In other words, $\rho$ represents a skewness between the
timelike or spacelike current distribution on cosmic string loops: if
$\rho=\rho_\text{\st{bias}}$, each loop contains an equal proportion
of timelike and spacelike contributions, while if
$\rho<\rho_\text{\st{bias}}$ (respectively
$\rho>\rho_\text{\st{bias}}$), the timelike (resp. spacelike) current
loss is dominant.

Various intuitive arguments can be offered for why loop creation may
favour or disfavour current losses, and even why there could be a bias
for charge over current losses (or vice versa). While we have not yet
come to compelling conclusions about these complex processes, we do
offer general arguments at the beginning of Sec.~\ref{chop} which tend
to disfavour charge or current loss by loops (i.e. $g>1$ or
$b<0$). However, this dynamical mechanism remains to be tested
directly by numerical simulations which is ultimately required to
calibrate the free parameters in the CVOS model. For this reason, a
variety of possible forms of these parameter dependencies and their
specific values will be explored throughout this work. A key goal is
to investigate how sensitive the string network evolution is to their
influence.

\subsection{Linear charge leakage} \label{IID}

One may anticipate that there should exist an additional energy loss
mechanism for a network of superconducting cosmic strings, namely the
so-called charge leakage, discussed e.g. in
Refs.~\cite{SpergelPiranGoodman, VilenkinVachaspati,
  BlancoPilladoOlum, MiyamotoNakayama, IbeKobayashiNakayamaShirai}. It
takes into account string curvature and the possibility that the
current does not follow the string trajectory exactly and may leave
the string~\cite{IbeKobayashiNakayamaShirai}; by construction, there
cannot be any such leakage from a straight string. Other effects
  should also be considered, such as a background of high energy
  particles hitting the relativistic strings, thereby increasing the
  energy of the condensate particles until they can escape. Also the
  reconnection of two strings, will disrupt the currents flowing along
  uncorrelated string regions which need to adjust, again potentially
  allowing the affected charges or currents to move off of the
  strings.

In the VOS model, there is only one characteristic length scale,
namely the average comoving inter-string distance or correlation
length $\xi_\mathrm{c}$, which can thus also be identified with the
average conformal string curvature $R_\mathrm{c} \approx
\xi_\mathrm{c}$.  Using this, we assume that the charge loss can be
embedded in the model by means of an additional term in the current
amplitude dynamics of the form
\begin{equation}
\label{ChargeLeakage}
\frac{\dd Y}{\dd \tau}\bigg|_{\text{leakage}} =  
- A \frac{Y}{\xi_\mathrm{c}} =  - A \frac{Y}{L_\mathrm{c}
\sqrt{1+Y}}, 
\end{equation}
where $A$ is a positive definite constant, characterising the amount
of charge leakage. To show that this is indeed the appropriate form,
let us consider the energy loss due to leakage.  This has to be
proportional to the current contribution in the total energy, which is
$E-E_0$. It also ought to be inversely proportional to the average
curvature $R_\textsc{c}$.  Putting together the above considerations
results in
\begin{equation}
\label{ChargeLoss}
\frac{\dd E}{\dd \tau}\bigg|_{\text{leakage}} 
=  - A \frac{E-E_0}{\xi_\mathrm{c}},
\end{equation}
which can be rewritten, using the characteristic length $L_\mathrm{c}$
and Brownian assumption \eqref{PhysCharLength}, together with the
relation \eqref{Energies} between the bare and total energies, as
\begin{equation}
\label{ChargeLoss2}
\frac{\dd L_\mathrm{c}}{\dd \tau}\bigg|_{\text{leakage}} =
\frac{A}{2} \frac{Y}{(1+Y)^{3/2}}.
\end{equation}
Eqs.~\eqref{ChargeLeakage} and \eqref{ChargeLoss2}
are consistent with one another provided that
there is no bare string energy change
due to charge leakage
\begin{equation}
\label{NoChargeLoss}
\frac{\dd E_0}{\dd \tau}\bigg|_{\text{leakage}} = 0,
\end{equation}
which should be the case because the charge and current contribution
is separate from the bare string energy.

The full system of dynamical equations for the extended VOS model
\eqref{EqOfMotMacroLin} with the addition of charge leakage
\eqref{ChargeLoss2} then can be written as
\begin{subequations}
\label{EqOfMotMacroLin2}
\begin{align}
\label{EqOfMotMacroLinA2}
\dot{L}_\mathrm{c} = &\  \frac{ L_\mathrm{c} \left( v^2  + Y  \right)}{1 + Y}
\frac{\dot a}{a} 
+ \frac{g c v (1+Y) + A Y}{2 (1+Y)^{3/2}}, \\
\label{EqOfMotMacroLinB2} \dot{v}  = &\ \frac{(1-v^2)}{1 + Y}
\left[ \frac{k\left(1 - Y \right) }{ L_\mathrm{c} \sqrt{1 + Y}}
- 2 v \frac{\dot{a}}{a}  \right],  \\
\label{EqOfMotMacroLinC2} \dot{Y}  = &\ 2 Y \left( 
\frac{v k}{ L_\mathrm{c} \sqrt{1 + Y}} - \frac{\dot{a}}{a}
\right) -  \frac{A Y}{L_\mathrm{c} \sqrt{1+Y}} \nonumber \\
& - \frac{v}{L_\mathrm{c}} c \left( g - 1
\right) \sqrt{1+ Y}, \\
\dot K = &\ 2 K \left( \frac{v k}{L_\mathrm{c}
\sqrt{1+ Y}} - \frac{\dot{a}}{a} \right) - 
\frac{2 (1-2 \rho_\textsc{a}) A Y}{L_\mathrm{c} \sqrt{1+Y}} \nonumber \\
&- 2 \frac{v}{L_\mathrm{c}} c \left( g - 1
\right) \left(1-2\rho\right)\sqrt{1+ Y}.
\label{EqOfMotMacroLinD2}
\end{align}
\end{subequations}
These are the general equations describing the time evolution of a
current-carrying string network with a linear equation of state with
the phenomenology discussed above taken into account. This
  system is more complicated than the VOS model as it seemingly
  doubles the number of degrees of freedom and adds extra undetermined
  parameters. It should however be emphasised that it represents a
  tremendous simplification of the true system (see Ref.~\cite{MPRS}),
  which is a large set of strongly coupled non linear partial
  differential equations; an intractable problem can instead be solved
  on a laptop.  The CVOS approach appears to be the only feasible
  route currently available for gaining insight into the time
  evolution of current-carrying cosmic strings.

We shall now consider special cases for the parameter choices to
clarify the existence and stability of scaling solutions and their
plausible cosmological consequences.

\subsection{Time evolution}

Consider the full system of equations for the superconducting network
Eq.~\eqref{EqOfMotMacroLin2}.  The first thing one notices is that, if
the four phenomenological parameters $k,c, g$ and $\rho$ do not depend
on $K$, then Eq.~\eqref{EqOfMotMacroLinD2} decouples from the rest of
the system \eqref{EqOfMotMacroLin2}, that is, the chirality $K$ is
sourced by the other variables but without backreacting on them. Note
that in general, i.e.\ for an arbitrary equation of state $F(K)$,
couplings do exist that are proportional to $F'(K)$ and $F''(K)$, so
that the above statement is only strictly valid in the case of a
linear equation of state \eqref{FLinear} for which $F'(K)=-\frac12$
(with $\kappa_0\to 1$) and $F''(K)=0$, both independent of $K$. Given
this decoupling, we shall not consider the time evolution of $K$ until
later in Sec.~\ref{StabilityK}.

We also need to specify at this point the cosmological background
evolution, which we assume to be first radiation and then matter
dominated: we want to investigate power-law expansion rates, for which
the scale factor evolves as $a \propto \tau^n$, so that we merely
replace $\dot{a}/a \to n/\tau$, with $n$ being a constant. The most
relevant cosmological regimes are $n=1$ for radiation domination and
$n=2$ for matter domination. Furthermore, we will also explore
numerically the cosmological transition from radiation to matter
domination, relying on the exact solution for the scale factor
\cite{Mukhanov:1990me}
\begin{equation}
    a(\tau) = a_\mathrm{eq} \left[ 2 \left(\frac{\tau}{\tau_\mathrm{eq}} 
    \right) + \left(\frac{\tau}{\tau_\mathrm{eq}} \right)^2 \right],
\end{equation}
which has the limits $a_\mathrm{rad} = 2
(a_\mathrm{eq}/\tau_\mathrm{eq}) \tau \propto \tau$ in the radiation
era, and $a_\mathrm{mat} = (a_\mathrm{eq}/\tau^2_\mathrm{eq})
\tau^2\propto \tau^2$ in the matter era, as required. The power
$$n = \displaystyle\frac{1+\left(\tau/\tau_\mathrm{eq}\right)}
{1+ \frac12 \left(\tau/\tau_\mathrm{eq}\right)}$$
indeed connects $n=1$ for $\tau\ll \tau_\mathrm{eq}$ to $n=2$
for $\tau\gg \tau_\mathrm{eq}$.

A scaling solution is characterized by $L_\mathrm{c}$ being a constant
fraction of $\tau$, so one can conveniently set $L_\mathrm{c} = \zeta
\tau$ to find
\begin{subequations}
\label{EqGenLin3}
\begin{align}
\label{EqGenLinA3}
\dot{\zeta} \tau = &\  \frac{v^2  + Y}{1 + Y} n \zeta
+ \frac{g c v (1+Y) + A Y}{2 (1+Y)^{3/2}} - \zeta, \\
\label{EqGenLinB3} \dot{v} \tau  = &\
\frac{1-v^2}{1 + Y} \left[
\frac{k\left(1 - Y \right)}{ \zeta \sqrt{1 + Y}}
- 2 n v \right],  \\
\label{EqGenLinC3} \dot{Y} \tau = &\ 2 Y \left(
\frac{v k }{\zeta \sqrt{1 + Y}} - n
\right) - \frac{v c (g-1) }{\zeta} \sqrt{ 1 + Y}\nonumber \\
& - \frac{AY}{\zeta\sqrt{1+Y}},
\end{align}
\end{subequations}
which contains all the modifications included in
Eq.~\eqref{EqOfMotMacroLin2} and should thus represent a reasonable
approximation of the dynamical evolution of the current-carrying
string network.

\section{Scaling solutions}
\label{II}

We start our study of current-carrying string network evolution by
making the simplest assumption that all the parameters described in
Sec.~\ref{IB} are constants; we then consider small linear deviations
of these from the standard values of the VOS model. Our underlying
assumption is that the linear equation of state \eqref{FLinear}, and
the corresponding Eqs~\eqref{EqOfMotMacroLin} and
\eqref{EqOfMotMacroLin2} on which the present analysis rests, hold for
small currents and therefore for a nearly Nambu-Goto string
network. Although, in principle, this corresponds to the small $K$
limit, we shall assume in what follows that $Y$ is also expected to be
small.

As discussed already above, one will also have to take into account
the fact that these phenomenological parameters can have a charge
dependence, which can play a role in the stability (or lack thereof)
of scaling solutions. The system \eqref{EqOfMotMacroLin2} decouples
$K$ from the other variables $L_\textsc{c}$, $v$ and $Y$ provided
these phenomenological parameters do not depend on $K$. Given our lack
of a better understanding of the microphysics involved, we shall
initially assume this below, i.e.\ $g=g(Y)$ and $c=c(Y)$, returning to
discuss the dynamics with possible chirality dependence in
Sect.~\ref{Chirality}.

Figure \ref{Fig:evol} provides an overview of this section by
displaying the possible cosmological evolution scenarios under various
modeling assumptions, while the network passes from the radiation to
matter dominated epochs. In what follows we discuss each of these
constant parameter scenarios in more detail, while noting that a
richer phenomenology emerges in Sec.~\ref{Seckv} using full numerical
solutions with variable parameters.

\subsection{Constant parameters}

We begin our analysis with a cosmological background dominated by a
single fluid component for which $n$ is constant, and we also neglect
charge and current leakage at the outset, thereby setting $A\to 0$.  A
scaling solution is one satisfying
\begin{equation}
\label{StdSacling}
v\to v\sca , \qquad \hbox{and} \qquad L_\mathrm{c} \to
\zeta\sca  \tau
\end{equation}
where $v\sca $ and $\zeta\sca $ are constants\footnote{A variable $X$
  at scaling will always be denoted by the same symbol with the
  scaling subscript, namely $X\sca$.  Although these are constants,
  they are to be contrasted with constant parameters entering the
  dynamical equations, denoted with a subscript ``o'' as in
  \eqref{ParamConst}.}.  Plugging the behaviours \eqref{StdSacling}
into the system of Eq.~\eqref{EqOfMotMacroLin2}, one notices that in
such a scaling regime the variables $Y$ and $K$ should also
asymptotically approach some constant values $Y\sca $ and $K\sca $. We
are thus led to find the equilibrium points for the system
\begin{subequations}
\label{EqGenLinA0}
\begin{align}
\label{EqGenLinAA0}
\dot{\zeta} \tau = &\  \frac{v^2  + Y}{1 + Y} n \zeta
+ \frac{g c v}{2 \sqrt{1+Y}} - \zeta, \\
\label{EqGenLinAB0} \dot{v} \tau  = &\
\frac{1-v^2}{1 + Y} \left[
\frac{k\left(1 - Y \right)}{ \zeta \sqrt{1 + Y}}
- 2 v n  \right],  \\
\label{EqGenLinAC0} \dot{Y} \tau = &\ 2 Y \left(
\frac{v k}{\zeta \sqrt{1 + Y}} - n
\right) - \frac{v c (g-1) }{\zeta} \sqrt{ 1 + Y}, \\
\dot{K} \tau = &\ 2 K \left(
\frac{v k}{\zeta \sqrt{1 + Y}} - n
\right)\nonumber\\
\label{EqGenLinAD0} &
- \frac{2 v c (g-1) }{\zeta} (1-2\rho) \sqrt{ 1 + Y}.
\end{align}
\end{subequations}

One way to obtain some constraints on the relevant parameter space is
to first assume that, at scaling, i.e.\ at the point (if any) for
which $\dot\zeta = \dot v = \dot Y = \dot K = 0$, the parameters of
the system \eqref{EqOfMotMacroLin} have reached constant values,
depending on the scaling values $\zeta\sca $, $v\sca $, $Y\sca $ and
$K\sca $, and so one sets
\begin{equation}
\label{ParamConst}
c\to c_o, \quad  k \to k_o, \quad g\to g_o, \quad \hbox{and} \quad 
\rho \to \rho_o,
\end{equation}
where $c_o$, $k_o$, $g_o$ and $\rho_o$
are constant. This is a natural requirement
as in any case they are expected to depend either
on the usual variables $\zeta$ or $v$, or on the
new $Y$ and $K$, all of which should behave as
constants at scaling.

Substituting the assumption \eqref{ParamConst} into
\eqref{EqGenLinA0}, and ignoring non-physical equilibrium points
(e.g., having $v\sca =1$ or $Y\sca ,\zeta\sca ,v\sca <0$), one obtains
the following values for the equilibrium point
\begin{equation}
\begin{aligned}
\label{ScalinLinGen}
v\sca^2 &= \frac{k_o}{n (c_o+k_o)} \frac{ k_o (n-2) +
c_o [2 (g_o - 1) + n]}{ k_o (n-2) + c_o (g_o-1+n)}, \\
\zeta\sca^2 &= \frac{k_o (c_o+k_o)}{4 n} \frac{k_o (n-2) +
c_o [2 (g_o - 1) +n]}{k_o (n-2) + c_o n}, \\
Y\sca &= \frac{ c_o (1 - g_o)}{ k_o (n-2) + c_o (g_o-1+n) }, \\
K\sca & =  \frac{ 2 c_o (1 - g_o) (1 - 2 \rho_o)}{  k_o (n-2)
+ c_o (g_o-1+n) },
\end{aligned}
\end{equation}
This includes the standard non-current-carrying VOS solution, having
$Y\sca = 0$ and $K\sca = 0$,
\begin{subequations}
\label{ScalinLinStd}
\begin{align}
v\sca ^2 &= v^2_\textsc{ng} \equiv \frac{k_o}{n (c_o+k_o)},
\label{ScalinLinStdv} \\
\zeta\sca ^2 &= \zeta^2_\textsc{ng} \equiv
\frac{k_o (c_o+k_o)}{4 n},\label{ScalinLinStdZ}
\end{align}
\end{subequations}
provided one assumes $g_o \to 1$, indicating, as expected, that the
model causes the cancellation of any averaged current if there is no
current chopping efficiency introduced.

It should be noted at this stage that Nambu-Goto and Abelian-Higgs
field theory network simulations
\cite{Moore,Martins:2005es,CorreiaMartins} have both shown
convincingly that not only does the relation \eqref{ScalinLinStd}
apply when $v\sca$ is seen as a constant obtained by solving the
implicit equation \eqref{ScalinLinStdv} with $k_o\to k(v\sca)$, but
also that\footnote{For the case of global (axion) strings
  \cite{Global} the values of the two parameters are less clear, due
  to the numerical difficulty of disentangling the effects of loop
  production and radiation losses.} for the cosmologically relevant
expansion rates (e.g., the radiation and matter eras) one can safely
assume that $c_o < k(v\sca)$, so that we can set $c_o<k_o$ when
investigating the solutions.  Moreover, the effect of charges or
currents is expected to decrease string velocities, thus increasing
the value of $k$. In this sense, the choice $k_o>c_o$ is physically
plausible and indeed using the parameter values obtained from NG and
Abelian-Higgs it is a conservative assumption.

\begin{figure*}
\begin{center}
\includegraphics[width=1.0\columnwidth,keepaspectratio]{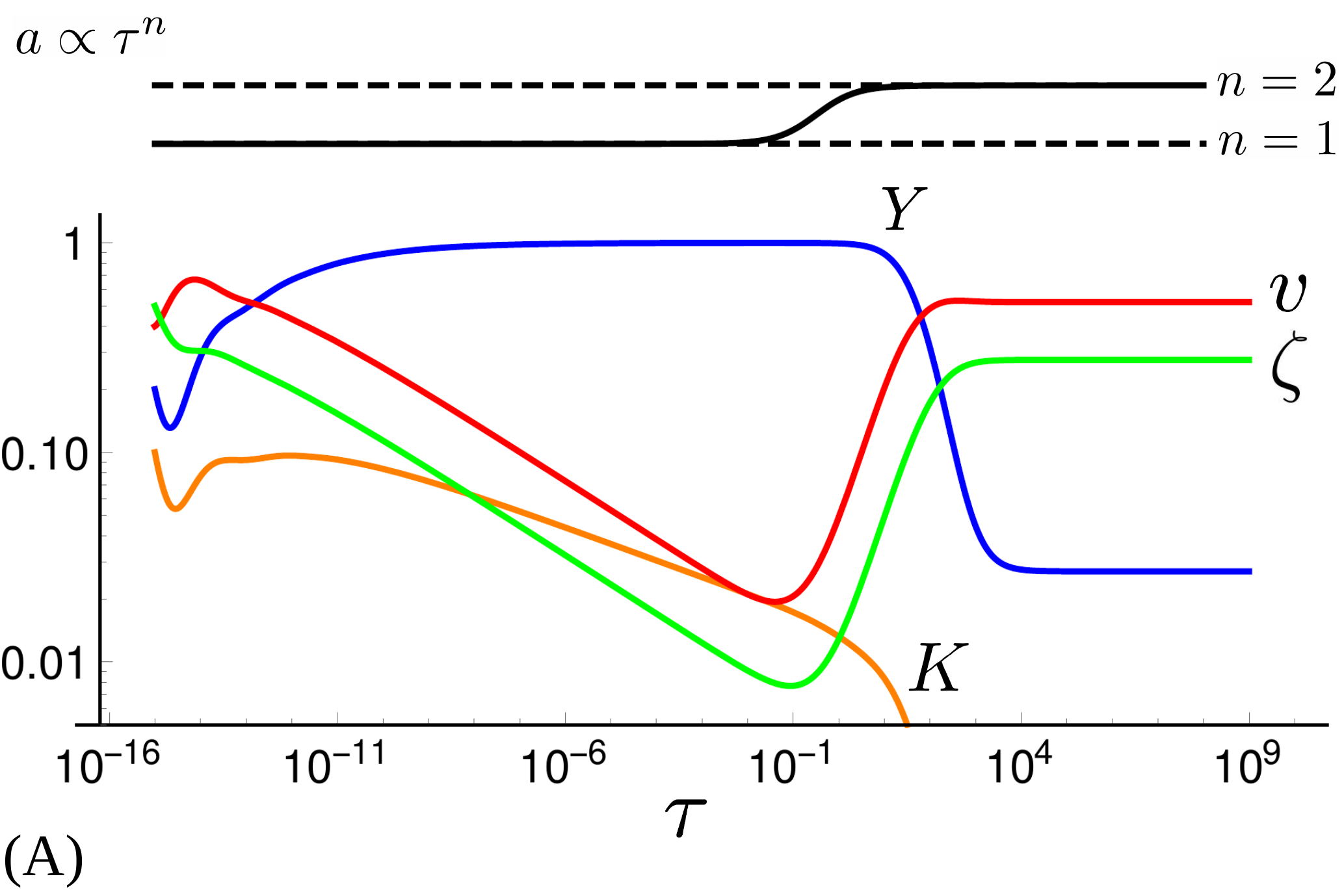}
\includegraphics[width=1.0\columnwidth,keepaspectratio]{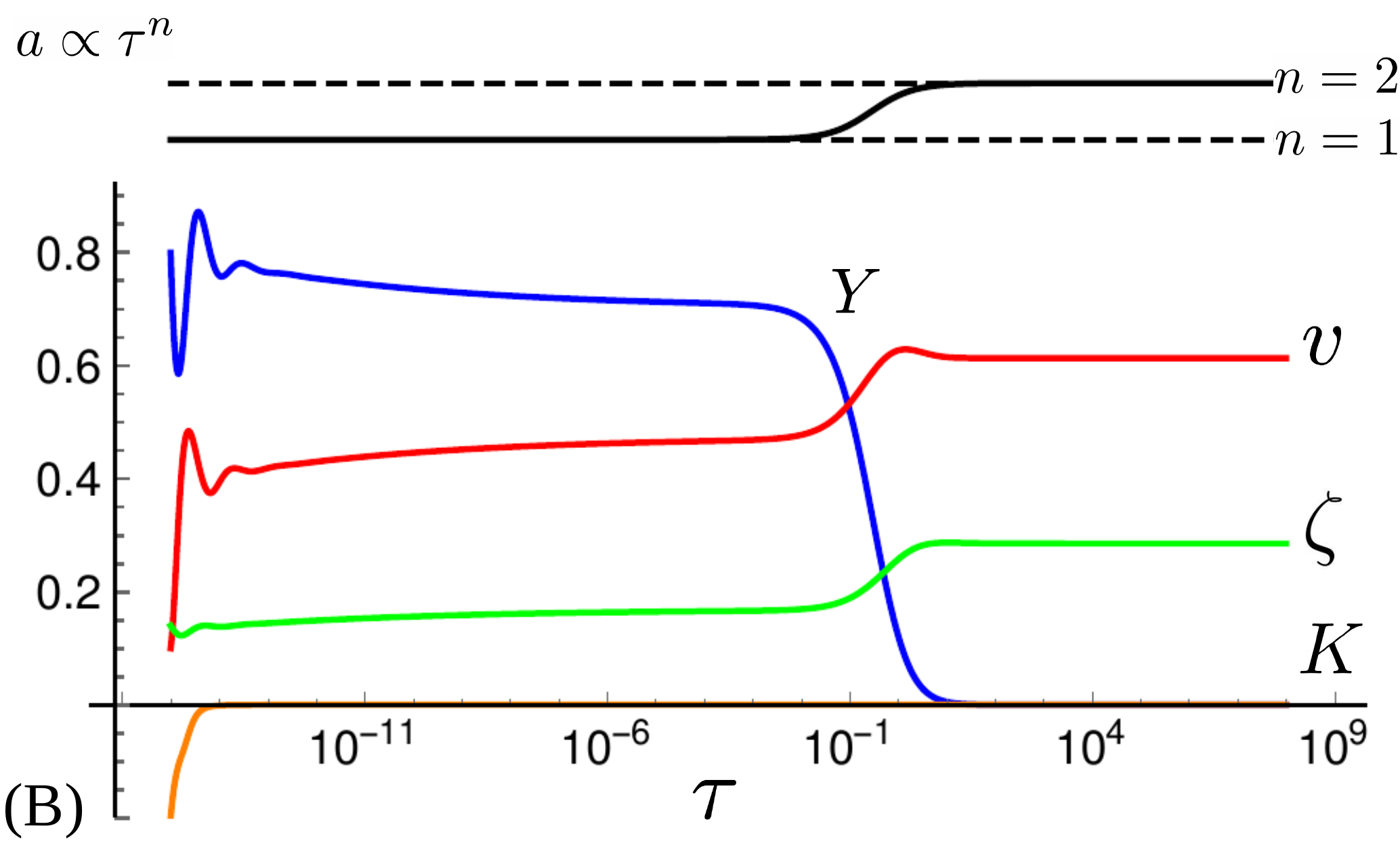}
\includegraphics[width=1.0\columnwidth,keepaspectratio]{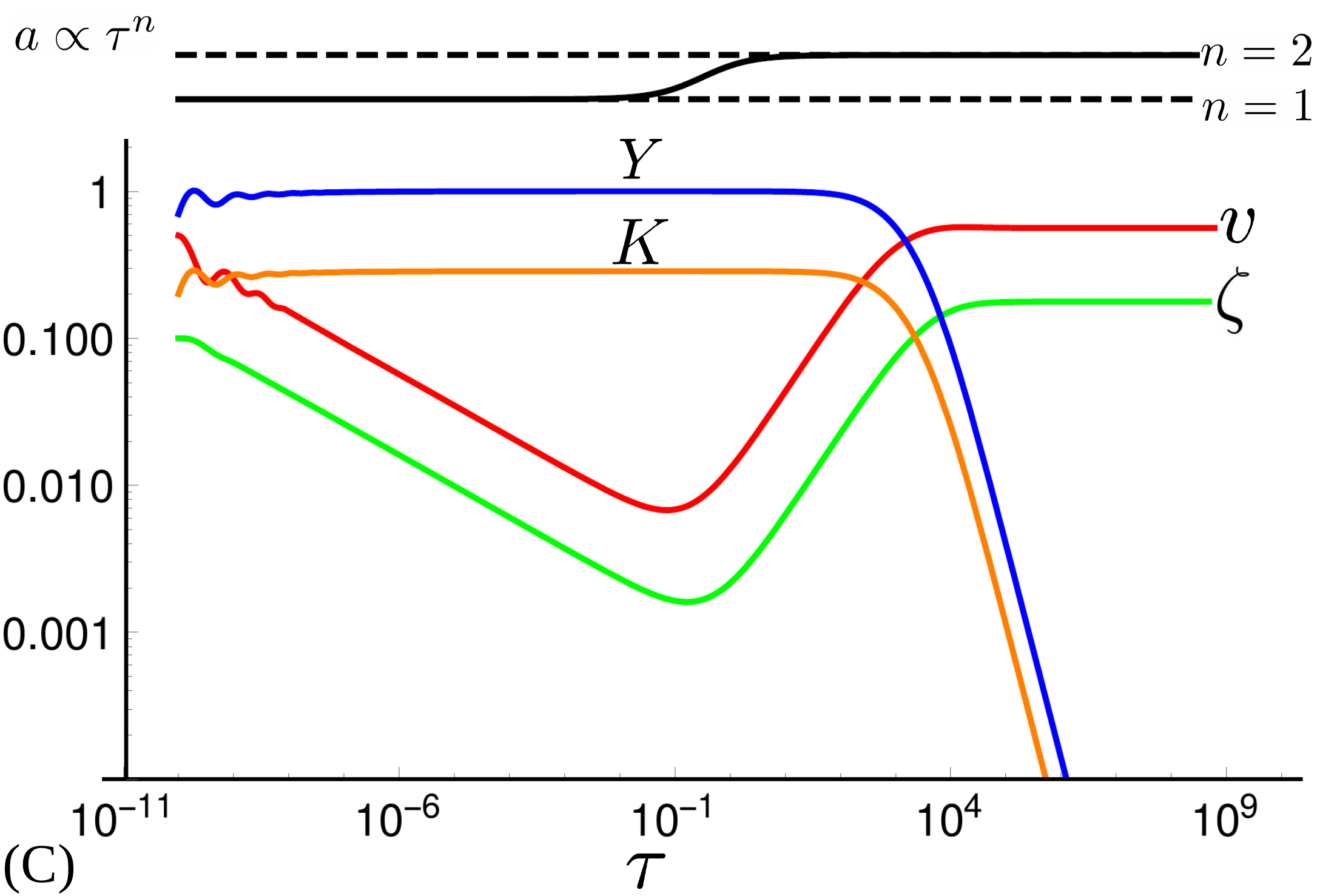}
\includegraphics[width=1.0\columnwidth,keepaspectratio]{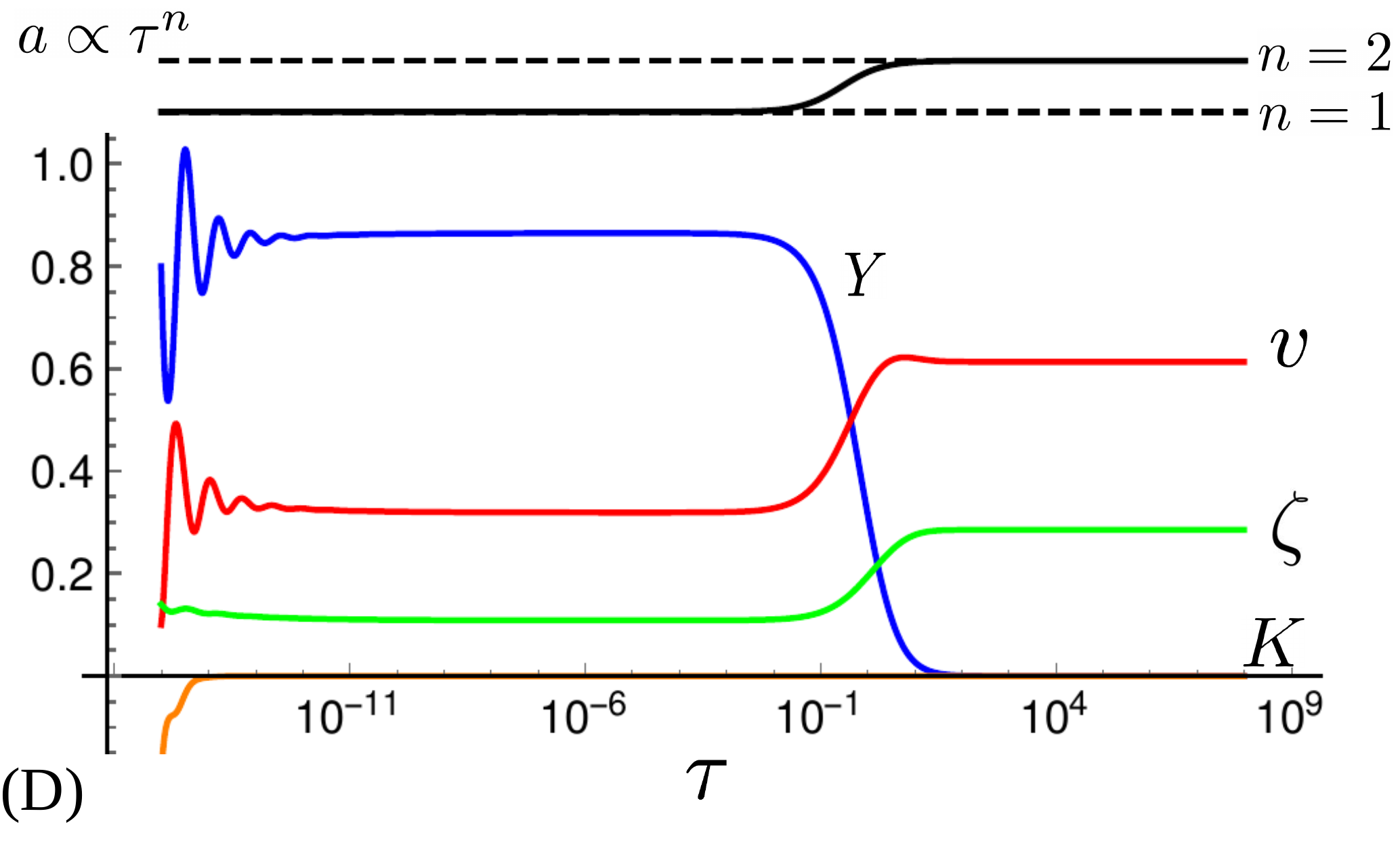}
\caption{\label{Fig:evol} Evolution of the velocity $v$, charge $Y$,
  chirality $K$ and characteristic length $\zeta$ in various cases
  through the radiation-to-matter transition (with $\tau_\text{eq}=1$,
  the time evolution of the expansion rate $n$ being shown above the
  graphs).\\ Top-left panel (A): solution of Eq.~\eqref{EqGenLinA0}
  with constant parameter values set to $g_o=0.9$, $c_o=0.5$,
  $k_o=0.6$ (with $\rho=1/2$ throughout), showing a non-scaling
  growing charge solution with $Y\to1$, $v\to0$ and $\zeta\to0$ during
  the radiation era, followed by a scaling charged configuration in
  the matter era.\\ Top-right panel (B): solution for
  Eqs.~\eqref{EqOfMotMacroLin3} with parameters set to $c_o=0.23$,
  $k_o=0.7$ and $b=0.6$, where a charged scaling solution in the
  radiation era evolves into a NG network in the matter
  era.\\ Bottom-left panel (C): solution with $c_o=0.23$, $k_o=0.4$
  and $g_o=1$, parameters for which the running solution of the
  radiation epoch, whose power laws are seen to satisfy
  Eq.~\eqref{NonStdScalinA}, evolves into a NG solution in the matter
  era.\\ Bottom-right panel (D): solution for
  Eqs.~\eqref{EqOfMotMacroLin5} with $c_o=0.23$, $k_o=0.7$, $b=0$ and
  $A=0.6$, in which case the growing charge solution is modulated into
  a charged scaling configuration in the radiation era, which then
  ends with an uncharged NG scaling solution in the matter era.  }
\end{center}
\end{figure*}

Let us begin with the radiation era, which, as we shall see below, is
more susceptible to exhibiting a non-trivial charged solution. In this
case, one finds a scaling charge
$$Y\sca^\text{rad} = \frac{c_o(1-g_o)}{c_o g_o - k_o},$$ together
with the other parameters $$v\sca^\text{rad} =
v^\text{rad}_\textsc{ng} \sqrt{1-Y\sca^\text{rad}}$$ and
$$\zeta\sca^\text{rad} = \zeta^\text{rad}_\textsc{ng}
\sqrt{1+\frac{2 c_o(1-g_o)}{k_o-c_o}}.$$ Provided $0\leq
Y\sca^\text{rad}<1$, this implies a slower moving network than
the NG case, $v\sca^\text{rad} \leq v^\text{rad}_\textsc{ng}$, as
expected. If $g_o>1$, the network is denser, with
$\zeta\sca^\text{rad} \geq \zeta^\text{rad}_\textsc{ng}$, and one
needs to ensure that $c_o g_o>k_o$, i.e. so there may exist a non
trivial scaling solution if the momentum parameter and the charge
chopping efficiency satisfy $c_o g_o>k_o>c_o$. On the other hand,
$g_o<1$ demands that $c_o g_o>k_o$ in order for
$Y\sca^\text{rad}$ to be positive; but $g_o<1$ also implies $c_o
g_o<c_o<k_o$, in contradiction with our hypothesis. With these
assumptions, the case $g_o<1$ generically then leads to a
non-scaling growing charge solution that we will discuss further
below.  In other words, if the usual NG relation
$k_o>c_o$ holds, the solution can only yield a scaling solution
with a charged network configuration, if $g_o>1$ in the radiation
era.

For the matter era, i.e. setting $n=2$, Eq.~\eqref{ScalinLinGen}
implies $Y\sca^\text{mat} = (1-g_o)/(1+g_o)$, and therefore a non
trivial scaling solution having $Y\sca>0$ requires $g_o<1$, regardless
of $c_o$ and $k_o$. Hence, a charged solution in the matter era with
$g_o<1$ implies a non-scaling growing charge solution in the radiation
era $Y\sca^\text{rad}\rightarrow 1$, as shown in Figure
\ref{Fig:evol}.A.  On the other hand, a charged scaling solution in
the radiation era implies an uncharged NG scaling solution in the
matter era, i.e. $Y\sca^\text{rad}\neq 0 \Longrightarrow
Y\sca^\text{mat}=0$ (similar to Figure \ref{Fig:evol}.B).

As final point it is worth clarifying the nature of the non-scaling
growing charge solution: in Eq.\eqref{EqGenLinAB0}, one sees that if
$Y(\tau)$ approaches unity, then the first term becomes negligible and
the dynamics is driven by the expansion only. The RMS velocity
$v(\tau)$ in that case decays, and so does $\zeta(\tau)$.  An example
of this behaviour is illustrated during the radiation era in
Fig.~\ref{Fig:evol} (on the left). Indeed, assuming power-law
behaviours for all the variables and taking the limit $\tau \to
\infty$, the leading terms in the CVOS model equations
\eqref{EqGenLinA0} lead to a growing charge solution of the form
\begin{equation}
\label{NonStdScalinA}
v \sim \tau^{-\alpha}, \quad
\zeta \sim \tau^{-\alpha} \quad \text{and} \quad
Y \sim 1 - \tau^{-2 \alpha},
\end{equation}
where the power $\alpha$ reads
\begin{equation}
\label{GrowAlpha}
    \alpha = 1 - \frac{n}{2} \frac{k_o+c_o (2g_o-1)}{k_o + c_o (g_o-1)}
\underset{g_o=1}{\to}
    1 - \frac{c_o+k_o}{2 k_o} n.
\end{equation}
This non-scaling growing or running solution apparently yields a
`frozen' network with maximum charge $Y\sim 1$; a comparison with
\eqref{Energies} then shows that this means the string network energy
is equally distributed between the bare string contribution and that
due to the charge. As this breaks our assumption of small currents
(the regime in which the linear equation of state should be valid), we
do not consider such solutions to be necessarily physically realised,
even if they exist mathematically.  Although this effect might
possibly be an artefact of the linear approximation, similar running
solutions should also be present for better motivated, nonlinear
equations of state~\cite{MPRS}. Most probably these solutions will not
come to a standstill, but rather lose charge and current through
microphysical effects, i.e.~achieving scaling through direct charge
leakage as we shall discuss in Sec.~\ref{leak}.

\begin{figure}
\begin{center}
\includegraphics[scale=0.56]{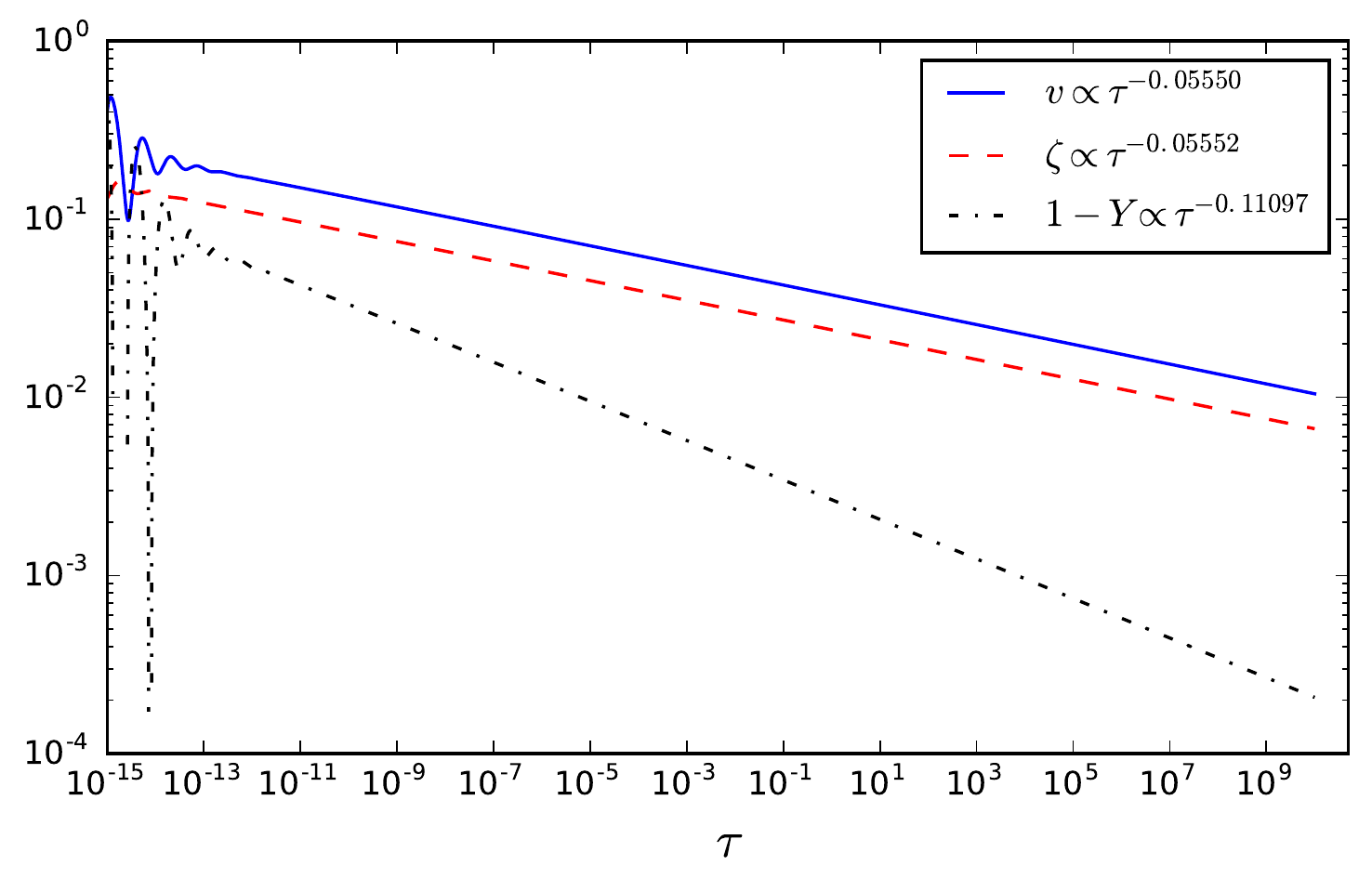}
\caption{\label{Figure:GrowingY}Evolution of
  Eqs.~\eqref{EqOfMotMacroLin5} with $c_o=0.8$, $n=1$, $A=10^{-7}$,
  $g=1$ and momentum parameter $k_o\to k(v)$ defined as
  \eqref{MomentumFunct} and comparison with the power-law decrease
  \eqref{NonStdScalinA}. The fitted values for the decay are obtained
  from the slope with $\tau>1$. The analytic value for the decay
  exponent is given by Eq.~\eqref{GrowAlpha}, numerically given by the
  value $0.05571$, where we used that $k(0) = 2 \sqrt{2}/\pi$.}
\end{center}
\end{figure}

Comparison of the numerical evolution of \eqref{EqGenLinA0} for $g=1$
with the analytic expression \eqref{NonStdScalinA} for $\alpha$ given
by \eqref{GrowAlpha} is shown in Fig.~\ref{Figure:GrowingY} and found
to be in good agreement with numerical calculations.  In practise, one
solves Eq.~\eqref{EqOfMotMacroLin5} in the limit of vanishingly small
leakage $A$ (see Sec.~\ref{leak} below): fixing $A\to 0$ exactly leads
to numerical issues as the time required to reach this asymptotic
behaviour scales inversely to $A$, and we have set $A\to 10^{-7}$.  To
be consistent with Sec.~\ref{Seckv}, we also fixed $k_o \approx k(0) =
2 \sqrt{2}/\pi$ in \eqref{GrowAlpha} [see Eq.~\eqref{MomentumFunct}
  for vanishing velocity].  It is interesting to note that a similar
effect was found in Ref.~\cite{Oliveira:2012nj} for a fixed momentum
parameter $k_o$.  Our growing charge solution with dynamical $k(v)$ is
illustrated in Fig.~\ref{Figure:GrowingY}. This explains the rather
counter-intuitive observation that some charge leakage is required to
produce a non-trivial scaling solution for a current-carrying network,
because charge growth needs to be modulated in some way, if dilution
due to the expansion or loop charge losses are inadequate.

\subsection{Current chopping bias}
\label{gbias}

Before moving to charge leakage, we consider non-constant values for
the current chopping efficiency parameter $g$, and in particular, we
ascribe a linear behaviour to $g$ as a function of the current
amplitude,
\begin{equation}
\label{GLinear}
g \equiv 1 + 2 b Y,
\end{equation}
where $b$ is a constant. This linear dependence is motivated by the
notion that the charges or currents can only have a significant
influence on the nature of loop production if they are non-zero.  In
order to clarify each effect separately, we assume, in this section,
that both the loop chopping efficiency $c$ and the momentum parameter
$k$ remain constant, and thus set $c\to c_o$ and $k\to k_o$.

Because of the decoupling of $K$ discussed above, we begin with the
sub-system
\begin{subequations}
\label{EqOfMotMacroLin3}
\begin{align}
\label{EqOfMotMacroLinA3}
\dot{\zeta} \tau = &\  \frac{v^2  + Y}{1 + Y}n \zeta
+ \frac{1+2 b Y}{2 \sqrt{ 1 + Y }} c_o v - \zeta, \\
\label{EqOfMotMacroLinB3} \dot{v} \tau  = &\
\frac{(1-v^2)}{1 + Y} \left[
\frac{k_o\left(1 - Y \right)}{ \zeta \sqrt{1 + Y}}
- 2 v n  \right],  \\
\label{EqOfMotMacroLinC3} \dot{Y} \tau = &\ 2 Y \left( 
\frac{v k_o}{\zeta \sqrt{1 + Y}} - n
\right) - 2 c_o b \frac{v}{\zeta} Y \sqrt{ 1 + Y}\,.
\end{align}
\end{subequations}
It is obvious that $Y=0$ is a solution of these, and corresponds to
the Nambu-Goto solution. Let us seek a different solution and assume
that further scaling with non-zero charge exists, whose equilibrium
point ($\zeta\sca $,$v\sca $, $Y\sca $) is obtained as the solution of
$\dot\zeta = \dot v = \dot Y = 0$. Writing the corresponding algebraic
equations in the form of a linear system in $c_o$, $n$ and $k_o$, one
finds
\begin{subequations}
\label{EqOfMotMacroLinSol}
\begin{align}
\label{EqOfMotMacroLinSolc}
c_o &= \frac{2 \zeta\sca  \sqrt{1+Y\sca } (2v^2\sca 
+Y\sca -1) }{v\sca  \left\{2 v^2\sca  [1+b(1+Y\sca )]+Y\sca
-1\right\} } , \\
\label{EqOfMotMacroLinSoln} 
n &= \frac{2 b (1 - Y\sca ) (1+Y\sca )}{2 v^2\sca
[1+b(1+Y\sca )]+Y\sca -1} ,  \\
\label{EqOfMotMacroLinSolb} 
k_o &= \frac{4 b v\sca  (1+Y\sca )^{3/2} \zeta\sca }{2 v^2\sca
[1+b(1+Y\sca )]+Y\sca -1}.
\end{align}
\end{subequations}
This is of course not to be mistaken as some fine-tuning of the
underlying parameters $n$, $c_o$ and $k_o$, which are fixed by either
the cosmological evolution or the local string physics. Obtaining
these relations and demanding that these parameters are positive
definite is merely one way of determining relevant constraints on
these parameters as well as on the scaling solution $\zeta\sca$,
$v\sca$ and $Y\sca$.  It is also a way to show the limited
applicability of such solutions.

One immediately notes that in the limiting case for which $b$ $\to$
$0$, the solution Eq.~\eqref{EqOfMotMacroLinSol} yields
$c_o$~$=$~$2\zeta\sca \sqrt{1+Y\sca}/v\sca$ and $n=k_o=0$, which would
therefore only apply in a Minkowski universe, unless the determinant
vanishes, which requires that $v\sca ^2 = (1-Y\sca )/2$. This in turn
implies $Y\sca <1$, in accord with the assumption that the linear
model ought to be valid for small current amplitudes\footnote{The
  small-current limit for which the linear equation of state is
  supposed to be valid concerns the chirality $K$, so that there does
  not seem to be any actual constraint on the charge $Y$. However, for
  $Y\sim 1$, the charge contribution to the overall network energy
  \eqref{Energies} is comparable to the bare energy, and this entails
  that the linear regime is no longer appropriate.}.  We shall keep
this assumption in what follows.

Requiring $n$, $c_o$ and $k_o$ to be positive, one obtains two
possible situations, depending on both the sign of $b$ and the sign of
the denominator $2 v^2\sca [1+b(1+Y\sca )]+Y\sca -1$ in
Eq.~\eqref{EqOfMotMacroLinSol}.  Note that under the assumption
$0<Y\sca <1$ discussed above, this denominator is actually positive
definite if $b>0$, which must be the case, according to
\eqref{EqOfMotMacroLinSolb}, if we demand an expanding universe with
$n>0$. Then, according to \eqref{EqOfMotMacroLinSolc}, we are left
with the requirement that $2v^2\sca +Y\sca -1 >0$, leading finally to
\begin{equation}
\label{YBoundsLinear}
1 - 2 v\sca ^2 < Y\sca  < 1.
\end{equation}
From the above and Eq. \eqref{GLinear}, we see that in order to have a
non-trivial current-carrying solution, we require $g>1$ which, we
recall, implies that chopped off loops carry more charge than the
infinite strings. In other words, it would mean that the average
charge is mostly carried by the loops. One therefore expects that in
the absence of this bias the charge on the long strings would grow.
This can also be understood by our boundary conditions, discussed
around Eq.~(29) in Ref.~\cite{MPRS}, according to which the integral
of the spatial derivative of any quantity over the full string network
should vanish. Assuming the current to come from such a (phase)
gradient, its overall value over the network should thus be initially
vanishing. As a result, any leftover should come from loops being
chopped off from the network.

Finally, if $b<0$ and $2 v^2\sca [1+b(1+Y\sca)]+Y\sca-1<0$, one can
obtain a condition for the scaling value of the current amplitude as
\begin{equation}
\label{YBoundsLinear2}
Y\sca < 1 - 2 v\sca^2,
\end{equation}
to ensure $c_o>0$; note that \eqref{YBoundsLinear2} is slightly more
restrictive than the original assumption as $b<0$ also implies $2
v^2\sca [1+b(1+Y\sca)]+Y\sca-1< 2 v^2\sca+Y\sca-1$. This describes the
case for which chopped off loops carry less charge than the infinite
strings. Note also that $b<0$ implies, because of \eqref{GLinear},
that $g$ could vanish or even become negative, which is impossible as
it would mean the total energy \eqref{EnergyLoss} increases as the
network forms loops! Clearly, while such solutions are mathematically
allowed, they are physically unrealistic.  In other words, with the
scaling solution $Y\sca$, one must impose that $b$ satisfies
\begin{equation}
b > - \frac{1}{2 Y\sca} > -\frac12,
\label{b12}
\end{equation}
where the last inequality was obtained for the limiting case $Y\sca
\to 1$, which is the maximum possible value. Here, we note the caveat
that at high charge levels $Y\rightarrow 1$, we expect all our linear
approximations to require modification.

Let us now investigate how the system of Eqs.~\eqref{EqOfMotMacroLin3}
depends on the momentum $k_o$, chopping $c_o$ and $b$ parameters.  In
this admittedly oversimplified case ($k_o$ and $c_o$ constant and no
direct charge leakage $A=0$), the scaling values (equilibrium points)
for the system of differential equations \eqref{EqOfMotMacroLin3} can
be found analytically (again excluding unphysical or non-scaling
cases).  There are two equilibrium points, the first is again the
standard, non current-carrying, solution \eqref{ScalinLinStd}, while
the second is a new solution with non-trivial current, namely
\begin{eqnarray}
\label{SolutionNonSt} 
v\sca^2 &=& \displaystyle\frac{k_o [c_o (n+4b) - k_o (2-n)]}{2 b c_o (c_o+k_o) n},
\nonumber \\
\zeta\sca^2 &=&
\displaystyle\frac{k_o (c_o+k_o)[c_o (n+4b) - k_o (2-n)]}
{4 n [k_o (2-n) - c_o n] }  , \\
Y\sca &=& \displaystyle\frac{k_o (2-n) - c_o (n + 2 b)}{2 b c_o},\nonumber
\end{eqnarray}
which reproduces \eqref{ScalinLinStd} when setting $k_o (2-n) = c_o
(n+2b)$. When $b\to0$, the above scaling solution is obviously
problematic as it implies infinite velocity and charge. Since both
$v\to1$ and $Y\to 1$ are singular points, the dynamics prevents such
solutions from being reached and naturally leads either to the growing
charge solution, as shown by Eq.~\eqref{NonStdScalinA} and illustrated
in the radiation era in Fig.~\ref{Fig:evol}.C, or to the uncharged NG
solution. For small values of $b$, actually including $b=0$, the
latter solution is the only available, and there exists a threshold in
$b$ above which \eqref{SolutionNonSt} becomes acceptable. This is
somehow similar to a symmetry breaking mechanism.

In the special case of matter domination ($n=2$), one finds $Y\sca =
-(1+b)/b$, which requires $-1<b<0$. Adding the extra requirement that
$v\sca^2>0$ (or $\zeta\sca^2>0$, both constraints being at this stage
equivalent) further restrict the available domain as it implies
$-1<b<-\frac12$. So, in order to sustain a non-trivial
current-carrying scaling solution during the matter era, one would
need a significant loop chopping bias against charge losses to
counteract the dilution caused by the enhanced expansion rate.  We do
not expect such a scenario to be physically realistic.

For general values of $n<2$, one finds first that for $b=0$, the
system is singular: there is no fixed point to the system
\eqref{EqOfMotMacroLin3} unless one imposes $k_o=c_o$, a rather
meaningless fine-tuning. Even then, the only solution would be trivial
from the point of view of the current ($Y\sca=0$), while $\zeta\sca =
c_o/\sqrt{2n}$ and $v\sca = 1/\sqrt{2n}$.  Clearly, in this case the
physically realistic solution will be one where the charge grows.

For $b\not=0$, one finds the following two possible cases:
\begin{itemize}
\item[$\star$] $b>0 \Longrightarrow c_o(n+4b)> k_o (2-n) > c_o (n+2b)$,
\item[$\star$] $b<0 \Longrightarrow c_o(n+4b)< k_o (2-n) < c_o (n+2b)$,
\end{itemize}
which can be summed up though the definition of the constant
\begin{equation}
\begin{gathered}
\label{RestrBRelation}
B \equiv b \left[ \frac{k_o}{c_o} (2-n)-n \right]^{-1},
\end{gathered}
\end{equation}
and the constraints now read as $\frac{1}{4}<|B|<\frac{1}{2}$.  An
example of time evolution for such a network is shown in
Fig.~\ref{Fig:evol}.B.

Under these assumptions, we find that for an FLRW background with
expansion rate bigger than for the matter dominated epoch ($n>2$),
there cannot be any non-trivial scaling solution ($Y\sca \neq 0$) with
$b \geq 0$. In other words, for such fast expansion rates the
expansion alone is sufficient to force the decay of the charge and
lead to the NG solution. As we shall see below, this is exactly what
is found numerically, with current-carrying scaling solutions which
are present in the radiation era quickly decaying towards a NG regime
as matter begins to dominate and the expansion rate increases.  As for
the radiation epoch ($n=1$), the constraint $b\geq 0$ implies that one
must have $k_o \geq c_o$ to achieve scaling.

Fig.~\ref{Fig:evol}.C demonstrates an example of a growing charge
solution, which appears when the standard NG equilibrium point
\eqref{ScalinLinStd} is a repeller and the equilibrium point
associated with non-trivial charge \eqref{SolutionNonSt} lies outside
the physically meaningful region. In the case of the growing or
running solution, we deviate from the usual string network scaling
behaviour and reach another power law description, given by
Eq.~\eqref{NonStdScalinA}.

\subsection{Linear charge leakage}
\label{leak}

Let us now focus on the effect of charge leakage, i.e.
Eqs~\eqref{EqGenLin3} with a nonvanishing current loss parameter $A$,
while again keeping $k=k_o$ and $c=c_o$ constant, but now neglecting
the current chopping efficiency with $g=1$ ($b=0$).  The system now
reads
\begin{subequations}
\label{EqOfMotMacroLin5}
\begin{align}
\label{EqOfMotMacroLinA5}
\dot{\zeta} \tau = &\   \frac{\left( v^2  + Y  \right)}{1 + Y} n \zeta +
\frac{c_o v (1+Y) + A Y}{2 (1+Y)^{3/2}} - \zeta, \\
\label{EqOfMotMacroLinB5} \dot{v} \tau  = &\ \frac{(1-v^2)}{1 + Y} 
\left[ \frac{k_o \left(1 - Y \right)}{ \zeta \sqrt{1 + Y}} - 2 v n 
\right],  \\
\label{EqOfMotMacroLinC5} \dot{Y} \tau = &\ 2 Y \left(
\frac{v k_o}{\zeta \sqrt{1 + Y}} - n
\right) - \frac{A Y}{\zeta \sqrt{1+Y}}.
\end{align}
\end{subequations}

One can find the equilibrium points for this system of differential
equations analytically. One of these points is the standard uncharged
scaling configuration defined by Eq.~\eqref{ScalinLinStd}, while
another one has the form
\begin{eqnarray}
v\sca &=& \frac{A}{k_o(2-n)-c_o n}, \nonumber \\
\zeta\sca &=& \frac{c_o+k_o}{2\sqrt{1+Y\sca}}
v\sca.\label{SolutNonTrLeak} \\
Y\sca &=& 1 - \frac{A^2 (c_o+k_o) n}{k_o [k_o (n-2) + c_o n ]^2},
\nonumber
\end{eqnarray}
whose limit, when $A\to 0$, is the frozen network of
Eq.~\eqref{NonStdScalinA}, as expected. Similarly to the solution
\eqref{SolutionNonSt} above, fixing $A$ such that $Y\sca\to0$ and
plugging back into $v\sca$ and $\zeta\sca$, one recovers
\eqref{ScalinLinStd}.  A particular example of such a string network
evolution is shown in Fig.~\ref{Fig:evol}.D.

Again, one immediately notices that this non-trivial equilibrium
can only be reached provided the expansion rate is $n \leq 2$, as
larger $n$ necessarily implies $v\sca<0$. A similar behaviour has
been identified for wiggly strings \cite{Almeida}. More
precisely, for $n\leq 2$, one has $0\leq v\sca <1$, i.e.
\begin{equation}
\frac{k_o}{c_o}  \geq \frac{n}{2-n}\ \underset{n\to 1}{\longrightarrow} \ 1,
\end{equation}
and
\begin{equation}
A \leq (2-n) k_o -c_o n,
\end{equation}
showing that, as expected, too much charge leakage leads back to the
non-current-carrying case.  This agrees with the previous discussion
when there were net charge losses due to loops with $g>1\; (b>0)$.
Demanding that $Y\sca >0$ leads to
\begin{equation}
v\sca^2 \leq \frac{k_o}{n (c_o+k_o)},
\end{equation}
a condition which is obviously satisfied as long as $n\geq 0$.

\subsection{Loop chopping parameter $c$}
\label{chop}

Having introduced biased loop chopping and charge leakage mechanisms,
we briefly consider the possibility of a charge-dependent loop
chopping efficiency, i.e. a decrease (or increase) of the amount of
loops produced depending on the presence of a charge along the
string. Motivation for such a possibility stems from previous works on
superconducting loops
dynamics~\cite{MartinsShellard1998,CarterPeterGangui,MartinPeter,Oliveira:2012nj},
showing that a current flowing along a string can prevent loops from
collapsing. Indeed, there exists a special value for the loop radius,
$r_\textsc{v}$ say, for which the loop reaches an equilibrium state
whereby the contraction due to the local string tension is balanced by
the angular momentum made possible by the Lorentz symmetry breaking of
the worldsheet due to the very existence of a current, with
$r_\textsc{v}$ actually depending on the current amplitude.

Clearly, for such a scenario, loop production would be strongly
suppressed for loop sizes smaller than $r_\textsc{v}$: the chopping
efficiency should be greatly reduced for such loops. While the typical
size of loops produced by superconducting networks is not well known,
one may nevertheless expect that it depends on the charge, and in our
modeling approach this would lead to a charge-dependent loop chopping
efficiency.  Alternatively, one may reason by analogy with the case of
wiggly strings, where one expects that there will be more loops at
small scale in comparison with the standard description (see
Ref.~\cite{Martins:2014,Vieira:2016} for details).  Again, the
chopping efficiency could depend on the charge.

To lowest order, such an effect could be modeled by the following
effective chopping parameter
\begin{equation}
\label{ChopEff}
c = c_o (1 -\beta Y),
\end{equation}
where $c_o$ is the standard, Nambu-Goto, chopping efficiency and
$\beta$ is a constant: as one expects the chopping to be reduced as
the charge increases, one could reasonably assume $\beta>0$.  Under
the hypothesis \eqref{ChopEff}, the VOS model equations take the form
\begin{subequations}
\label{EqOfMotMacroLin7}
\begin{align}
\label{EqOfMotMacroLinA7}
\dot{\zeta} \tau = &\ \frac{v^2  + Y}{1 + Y} n\zeta +
\frac{c_o \left(1-\beta Y \right) v }{2 \sqrt{1+Y}} - \zeta, \\
\label{EqOfMotMacroLinB7} \dot{v} \tau  = &\ \frac{(1-v^2)}{1 + Y} \left[
\frac{k_o \left(1 - Y \right)}{ \zeta \sqrt{1 + Y}} - 2 v n  \right],  \\
\label{EqOfMotMacroLinC7} \dot{Y} \tau = &\ 2 Y \left( 
\frac{v k_o}{\zeta \sqrt{1 + Y}} - n
\right),
\end{align}
\end{subequations}
where all the previously discussed charge loss mechanisms are
neglected, i.e. we set $g=1$ and $A=0$.

Clearly, the standard NG behavior with $Y=0$ is a solution of this
system. On the other hand, if one enforces $Y\neq0$ one nominally
finds the non-trivial equilibrium point of
Eqs.~\eqref{EqOfMotMacroLin7} as
\begin{eqnarray}
v\sca^2 &=&
\frac{k_o (2-n) + \left( \beta -1\right) n c_o}
{2 \beta n c_o}, \nonumber \\
\zeta\sca^2 &=& \left( \frac{k_o^2}{2 n^2} \right)
\frac{k_o (2-n) +
\left( \beta-1\right) n c_o }{ k_o (n-2) +
\left( \beta+1\right) n c_o },
\label{SolutNonC} \\
Y\sca &=& \frac{ k_o (n-2) + n c_o}{ \beta n c_o}.\nonumber 
\end{eqnarray}
This solution shares with that obtained as \eqref{SolutionNonSt} that
it becomes unreachable for sufficiently small values of $\beta$ since
both the charge and velocity would diverge as $\beta\to0$, while in
the same limit $\zeta\sca^2$ would be negative. The same conclusion
actually holds that for small values of $\beta$, only the NG attractor
is dynamically attainable, and there exists a threshold in $\beta$
above which the new charged solution becomes physically admissible as
the symmetry broken phase for low enough temperatures. It remains to
be seen whether such a solution is realised in practice, i.e. is an
attractor or a repeller.

The specific case of the matter-dominated regime with $n=2$ yields the
special solution $Y\sca = 1/\beta$, which requires $\beta > 1$, in
order to ensure that $v\sca >0$ and $\zeta\sca^2>0$. Note at this
point that \eqref{ChopEff} would then imply that the effective
chopping parameter $c$ vanishes at scaling, while in the limit
$\beta\to \infty$, the RMS velocity becomes $v\sca\to
1/\sqrt{2}$. This is in agreement with the fact, known from both the
VOS model and from NG numerical simulations, that a non-zero loop
chopping efficiency is not necessary for a string network's density
and velocity to reach scaling: a fast enough expansion rate, including
the matter era, is sufficient \cite{Martins2016B}.

In summary, the outcome of this exercise is that for $n>2$ the NG
solution is the only possible one, while for $n<2$ we would have an
ill-defined solution if we insisted on a constant charge. Clearly in
this regime we should either expect a growing charge solution or will
need some leakage mechanism to ensure charge scaling for small
expansion rates.  In other words, a charge-dependent loop chopping
efficiency does not lead to any qualitatively new behaviour with
respect to what has been discussed in the previous sub-sections. So we
note the three different solution classes, depending on the expansion
rate--Nambu-Goto behaviour for fast expansion rates, growing charge
for small expansion rates (unless modulated by charge leakage), and a
transition between the two occurring at the matter-dominated era.
Similar behaviour has also been identified for wiggly cosmic strings
\cite{Almeida}. We present further supporting evidence for these
solutions in the sections that follow.

\section{Scaling stability}

In order to unveil the nature of the critical points described in the
previous section, we expand the relevant quantities around each of
these solutions.

\subsection{General method}
\label{Stability}

The existence of equilibrium points with non-trivial current, such as
those given by Eqs.~\eqref{SolutionNonSt}, \eqref{SolutNonTrLeak} and
\eqref{SolutNonC}, does not guarantee that these points can be
dynamically reached, and even less that they represent attractors for
the corresponding systems of differential equations. To understand the
nature of these, we study the relevant Jacobian matrices.  For
convenience we introduce the vector $\bm{V} = \left\lbrace \zeta, \,
v, \, Y \right\rbrace $ and the vector function $\bm{Z}(\bm{V})$,
which allows us to rewrite formally the systems
\eqref{EqOfMotMacroLin3}, \eqref{EqOfMotMacroLin5} and
\eqref{EqOfMotMacroLin7} as
\begin{equation}
\label{Pert1}
\dot{\bm{V}} \tau = \bm{Z}(\bm{V}).
\end{equation}

\begin{figure}
\begin{center}
\includegraphics[scale=0.52]{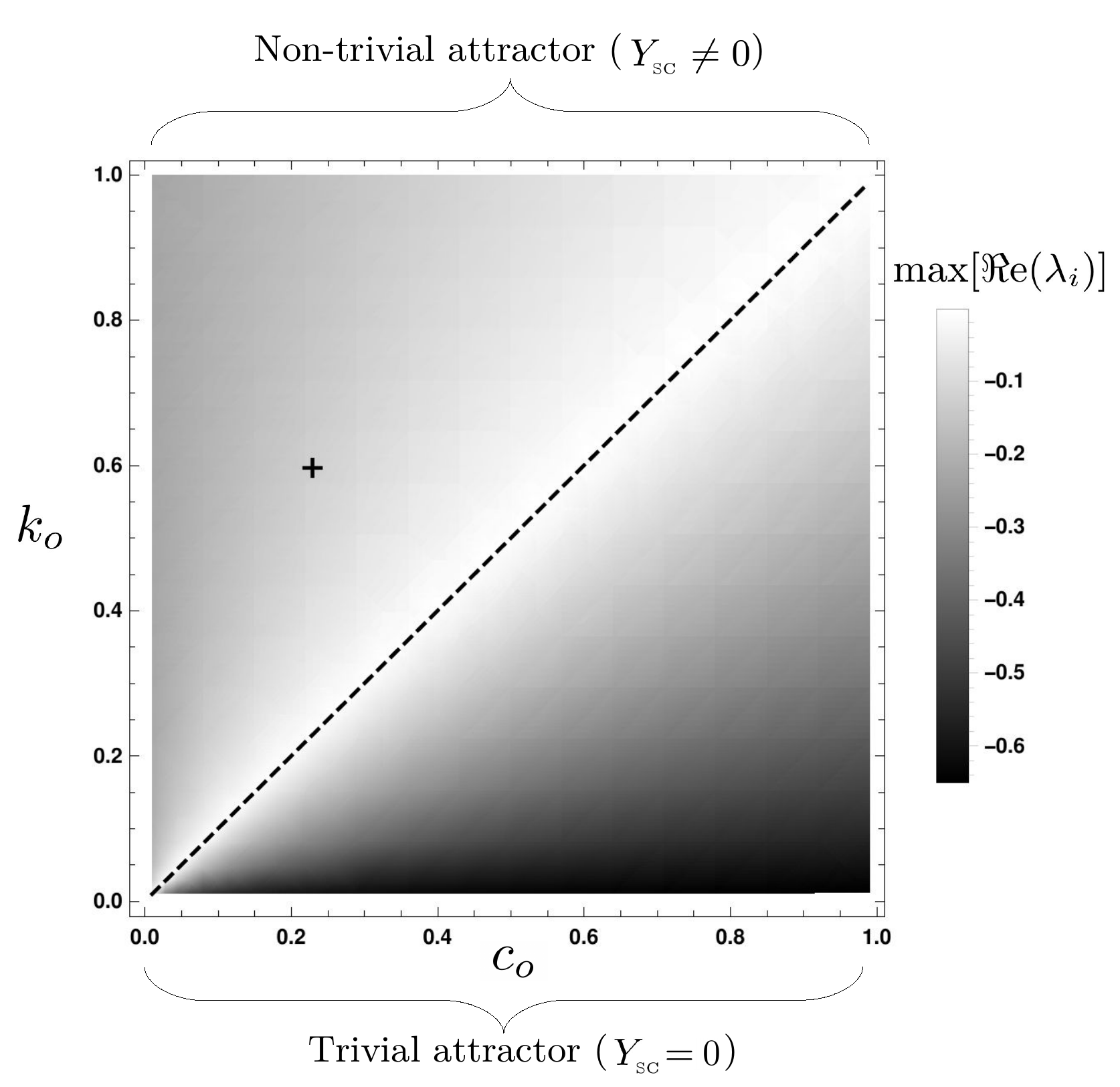}
\caption{\label{Figure:c-ko} Distribution of the maximal real part of
  the Jacobian eigenvalues \eqref{Pert3} around the no-current
  \eqref{ScalinLinStd} (lower right triangle) and current-carrying
  \eqref{SolutionNonSt} (upper left triangle) for $B=\frac13$. Note
  that changing the value of $B$ in the range allowed y the constraint
  below \eqref{RestrBRelation} does not qualitatively change the
  plot. The dashed line represents the values of $k_o$ and $c_o$ for
  which $\text{max}[\Rez(\lambda_i)]=0$, so that the upper left
  triangle represents the region for which the equilibrium point
  \eqref{SolutionNonSt} is an attractor, while \eqref{ScalinLinStd} is
  an attractor in the lower right triangle.  Varying the expansion
  rate $n$ changes the slope of the dashed line, which coincides with
  Eq.~\eqref{k_o-cLinear}.  An example of a trajectory for a non
  trivial solution is shown on Fig.~\ref{Figure:PhaseDiagrLin} for the
  parameter values at the point denoted by the $+$ sign ($c_o=0.23$
  and $k_o=0.6$).  }
\end{center}
\end{figure}

\begin{figure}
\begin{center}
\includegraphics[scale=0.58]{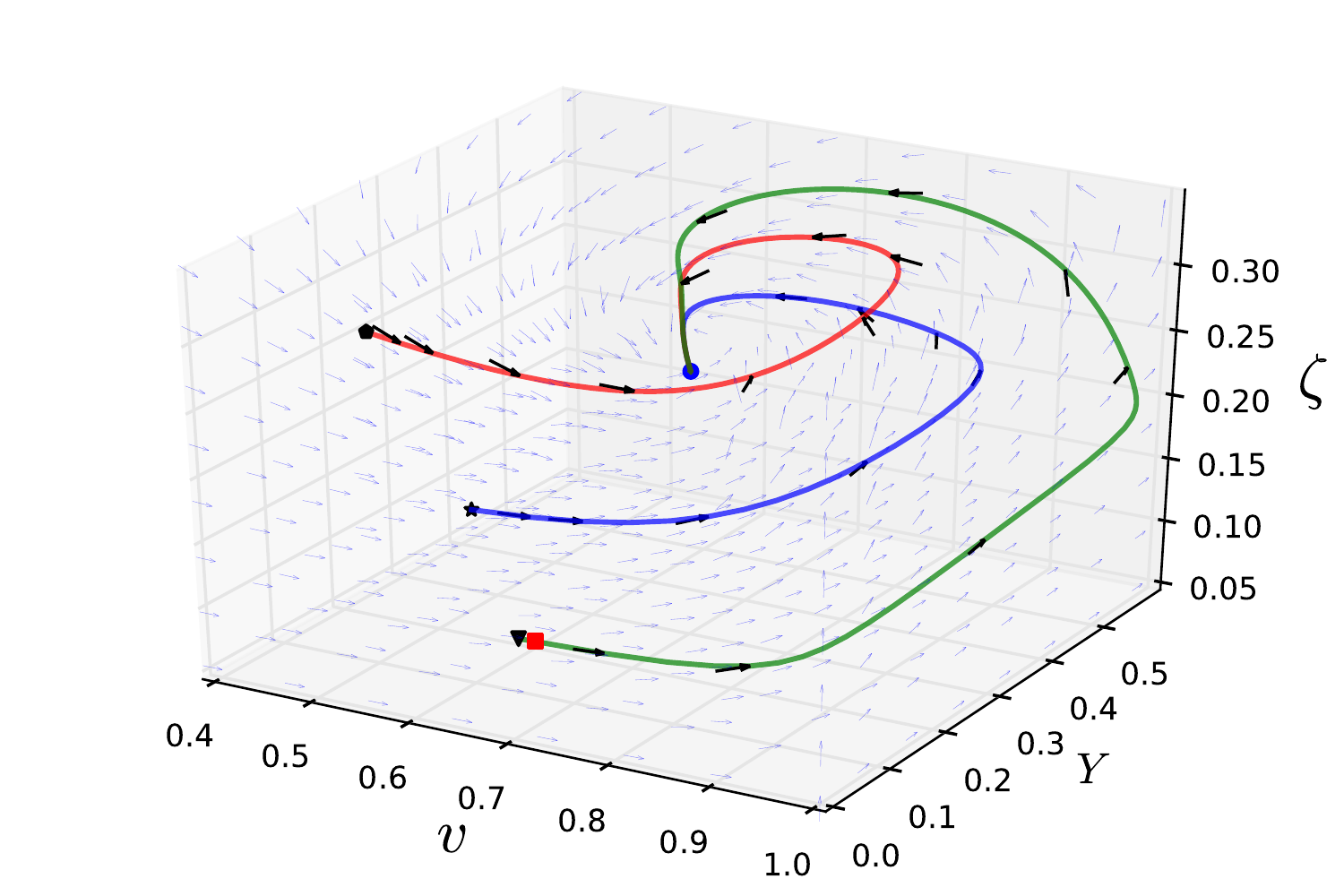}
\caption{\label{Figure:PhaseDiagrLin} Phase diagrams for the non
  trivial equilibrium points shown on Fig.~\ref{Figure:c-ko} with
  parameters $c_o=0.23$, $k_o=0.6$, $b=0.4$ and $n=1$
  (radiation-dominated epoch).  The non-trivial solution
  \eqref{SolutionNonSt} is an attractor in this case. It is shown by
  the blue round point, while the uncharged solution
  \eqref{ScalinLinStd}, a repeller, is represented as a red square
  point.  Different lines represent different evolutions (different
  initial conditions) of the system \eqref{EqOfMotMacroLin3}.  The
  gray dots represent initial conditions.}
\end{center}
\end{figure}

Denoting $\bm{V}\sca \equiv \left\lbrace \zeta\sca, \, v\sca, \, Y\sca
\right\rbrace$ the equilibrium (scaling) point, i.e. the solution of
$\bm{Z}(\bm{V}\sca) = 0$, the stability of $\bm{V}\sca$ can be
understood by finding the parameters for which small perturbations
around the equilibrium points decay with time.

Let us consider a solution $\bm{V}$ representing a small deviation
from the equilibrium point, namely with
$|\delta\bm{V}|\ll|\bm{V}\sca|$, where we defined $\delta \bm{V} =
\bm{V} - \bm{V}\sca$.  We expand the vector function $\bm{Z}(\bm{V})$
in \eqref{Pert1} as

\begin{equation}
\label{Pert2}
\dot{\bm{V}} \tau = \bm{Z}(\bm{V}\sca) +
\frac{\partial \bm{Z}(\bm{V})}{\partial \bm{V}}
\Bigg\vert_{\bm{V} = \bm{V}\sca} \delta
\bm{V} + \mathcal{O} \left( \delta \bm{V}^2 \right),
\end{equation}
which can be reduced to 
\begin{equation}
\label{Pert3}
\delta \dot{\bm{V}} \tau = \mathcal{J}\sca \delta \bm{V},
\end{equation}
where $\mathcal{J}\sca$ is the Jacobian matrix $\partial
\bm{Z}/\partial \bm{V}$ evaluated at the point $\bm{V} =
\bm{V}\sca$. Diagonalizing $\mathcal{J}\sca$ through $\mathcal{J}\sca
= \mathcal{P}^{-1} \mathcal{D} \mathcal{P}$, thereby defining the
matrix $\mathcal{P}$, and setting $\mathcal{D} = \text{diag} \left(
\lambda_1, \lambda_2, \lambda_3 \right)$, one can integrate
Eq.~\eqref{Pert3} to obtain the time dependence of the eigenvector
components as
\begin{equation}
\label{Pert4}
\left( \mathcal{P}\delta\bm{V}\right)_i \propto \tau^{\lambda_i}.
\end{equation}
The scaling solution $\bm{V}\sca$ is an attractor
if the perturbation $\delta\bm{V}$ decreases for
large $\tau$, which translates, through \eqref{Pert4}, into
a requirement on the real part of the eigenvalues of the
Jacobian $\mathcal{J}\sca$, namely $\Rez (\lambda_i)\leq 0,
\forall i$. In what follows, we study the distribution of the
maximum $\Rez (\lambda_i)$ as function of the relevant parameters
to determine the attractor regions.

\subsection{Current chopping bias}

The CVOS model, in the version described in Sec.~\ref{gbias} by
Eqs.~\eqref{EqOfMotMacroLin3}, contains four parameters, namely $k$,
$c$, $b$ and $n$, where $k$ is in principle a function of the RMS
velocity $v$, and $c$ and $b$ could depend on the charge, as discussed
in the previous sections.

While the expansion power index $n$ is fixed by the background
cosmological setup, the other quantities are in principle given by the
microphysics of the strings themselves. Lacking knowledge of their
actual numerical values, one must choose a set $\{c_o,k_o,b\}$ leading
to physically meaningful solutions for the scaling RMS velocity
$v\sca$, characteristic length $L\sca = \zeta\sca \tau$ and charge
magnitude $Y\sca$, given the restriction on $B$ as defined in
\eqref{RestrBRelation}.

Fig.~\ref{Figure:c-ko} illustrates our procedure for $n=1$ (radiation
domination epoch) by setting $B \to \frac13$ and plotting the maximum
value of the real part of the eigenvalues $\lambda_i$ in the
$(c_o,k_o)$ plane for both equilibrium points \eqref{ScalinLinStd}
(Nambu-Goto network) and \eqref{SolutionNonSt}.

Carrying out calculations for different values of the underlying
parameters, one notices that only the expansion rate $n$ and the ratio
$k_o/c_o$ are important to determine the nature (attractor, repeller,
saddle point) of the equilibrium point. Moreover, the parameter space
where the equilibrium point with the trivial magnitude of the current
$Y\sca=0$, given by Eq.~\eqref{ScalinLinStd}, is an attractor does not
overlap with the one where the equilibrium point with non-trivial
current magnitude $Y\sca \neq 0$, given by Eq.~\eqref{SolutionNonSt},
is an attractor. This behaviour means that, depending on the
phenomenological parameters, one expects one solution only to be
realised, either charged or uncharged.  In
Fig.~\ref{Figure:PhaseDiagrLin} an example is shown of the phase space
trajectory of the system \eqref{EqOfMotMacroLin3} in the radiation era
($n=1$) which approaches a non-trivial scaling charge magnitude; this
uses constant parameter values $c_o$ and $k_o$ consistent with a NG
scaling solution with some loop charge loss
$b>0$. Fig.~\ref{Figure:PhaseDiagrLin} also illustrates the
independence of the initial conditions for the charged attractor
solution under these assumptions.

The relevant regions of parameter space representing charged and
uncharged attractors are separated by the line
\begin{equation}
\label{k_o-cLinear}
k_o = \frac{n}{2 - n} c_o,
\end{equation}
implying in particular that for matter domination era ($n=2$),
the non-trivial charged scaling solution triangle apparently shrinks to
the line $c_o=0$. For $n>2$, as both $k_o$ and $c_o$ are
positive definite, there is no stable attractor solution
with non-trivial current. Similar behaviours for the existence
(or otherwise) of scaling solutions has also been observed for
chiral superconducting strings \cite{Oliveira:2012nj} and wiggly
strings \cite{Almeida}. Here, however, we reiterate the caveat
that these assumptions should not exclude the possibility of a
physical growing current solution discussed previously which
can be modulated by other charge loss mechanisms.

\subsection{Linear charge leakage}
\label{Linear charge leakage}

\begin{figure}
\begin{center}
\includegraphics[scale=0.52]{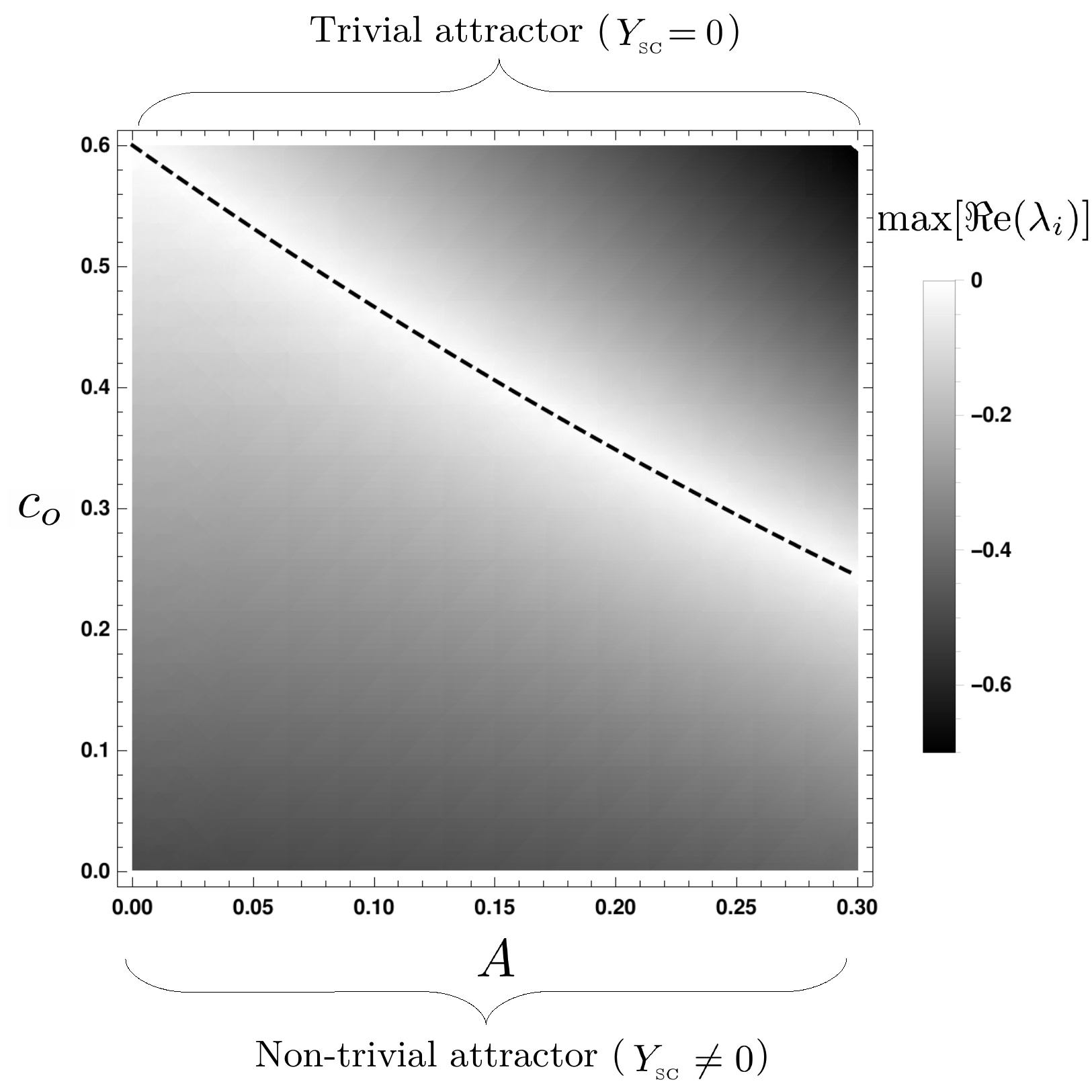}
\includegraphics[scale=0.52]{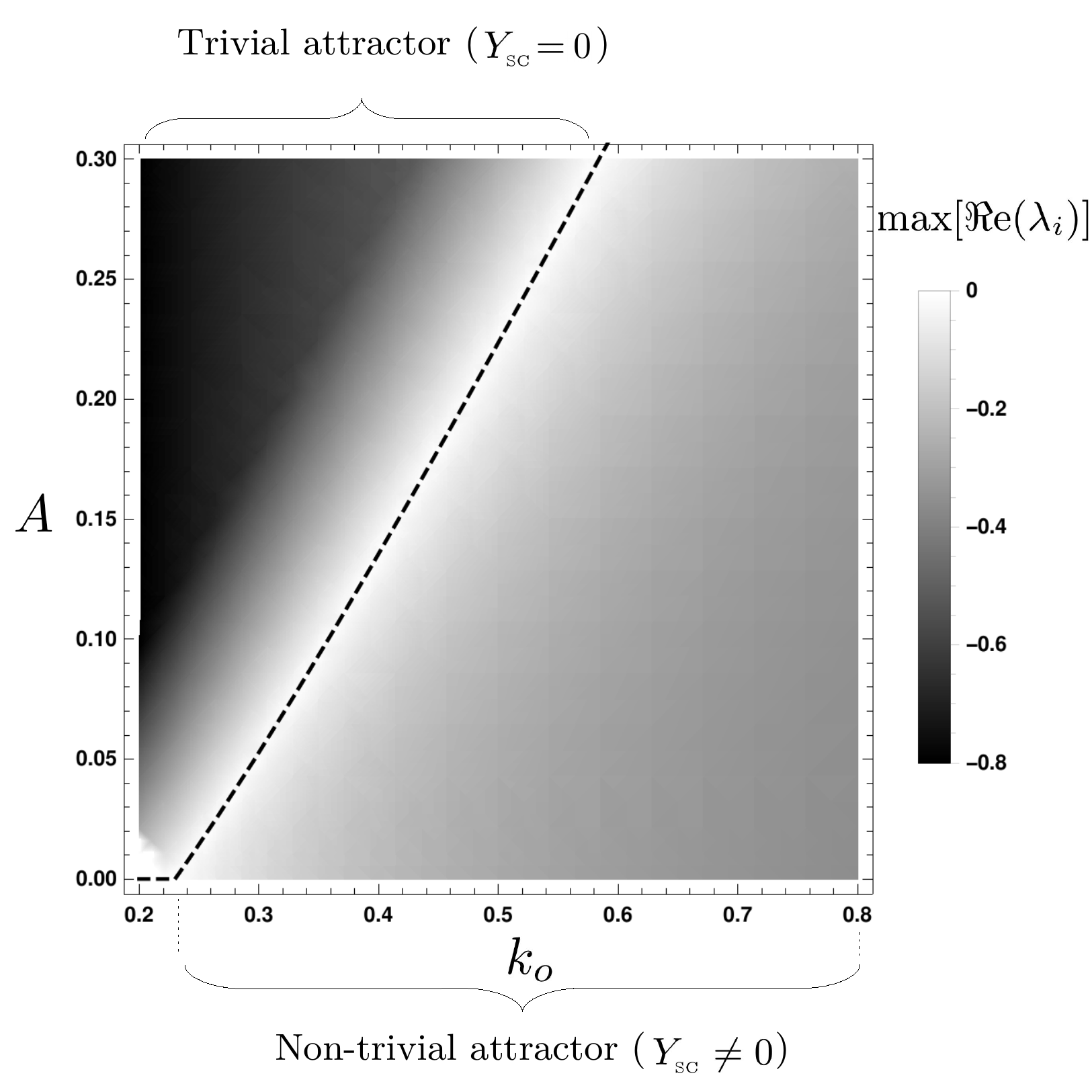}
\caption{\label{Figure:Leakage} Distribution of the maximal real part
  of the Jacobian eigenvalues \eqref{Pert3} for the system of
  \eqref{EqOfMotMacroLin5} with fixed points given by
  \eqref{ScalinLinStd} and \eqref{SolutNonTrLeak}, for $n=1$
  (radiation dominated era).  Only the regions with negative
  eigenvalues, representing attractors, are shown.\\ Upper panel: $(c,
  A)$ plane -- $k_o = 0.6$.\\ Lower panel: $(k_o, A)$ plane --
  $c_o=0.23$.\\ The fixed point \eqref{SolutNonTrLeak} with
  $Y\sca\not=0$ is a repeller (respectively an attractor) in the
  region of parameter space, where the trivial one,
  \eqref{ScalinLinStd}, with $Y\sca=0$, is an attractor (resp. a
  repeller). The dashed line that separates both regions is given by
  Eq.~\eqref{LineLeakage}.  }
\end{center}
\end{figure}

We now return to the linear leakage model \eqref{EqOfMotMacroLin5} of
Sec.~\ref{leak} with fixed points given by \eqref{ScalinLinStd} and
\eqref{SolutNonTrLeak}, to which we apply the eigenvalue method of
Sec.~\ref{Stability}.  We make the simplifying assumption again that
there is no loop charge bias $g=1$ (or $b=0$).  Since the parameter
space for this model is 3-dimensional, we chose to plot two slices of
the distribution of eigenvalues, namely in the $(A,c_o)$ and $(k_o,A)$
planes. Fig.~\ref{Figure:Leakage} shows the distribution of maximal
real part of the eigenvalues for the radiation epoch ($n=1$).

As we see from this figure, there are two non-overlapping regions of
parameter space, one with the uncharged solution attractor with
$Y\sca=0$, and the other with $Y\sca \neq 0$. It is worth emphasizing
at this point that the two solutions are actually exclusive of one
another, so the system is completely deterministic and mostly
independent of the initial conditions, the asymptotic solution
features being determined by the values of the underlying
parameters. Note also that there exist other mathematically acceptable
solutions with non-physical values of the variables, such as
e.g. having $Y\sca <0$ or $\zeta\sca<0$; we will discard these
solutions which can be found in the region for which the attractor is
for $Y\sca=0$.

Repeating the analysis of Fig.~\ref{Figure:Leakage} for different
values of $n$, one finds that the line that separates the two regions
of parameter space is described by the equation
\begin{equation}
\label{LineLeakage}
A^2 = \frac{k_o [k_o (2-n) - c_o n ]^2 }{(c_o+k_o) n},
\end{equation}
with the condition $k_o (2-n) - c_o n>0$.

\section{Chirality}
\label{StabilityK}

Up to this point, we have analyzed the stability of scaling solutions
for different phenomenological scenarios of the system of
Eqs.~(\ref{EqOfMotMacroLin2}) without the chirality $K$.  As we argued
already, provided the phenomenological parameters do not depend on $K$
and in the linear equation of state with which we are concerned here,
this variable is decoupled from the rest of the system and does not
contribute to the string network evolution.  Nevertheless one should
check that it also has a stable scaling solution.  If we start at the
current-forming phase transition at time $\tau_0$ with purely chiral
initial conditions $K(\tau_0)=0$, it is expected to remain zero
throughout the subsequent evolution of the network, but if we start
with arbitrary $K(\tau_0)$, we need to require that the scaling value
$K\sca$ is an attractor.  Physically, as the initial current exists as
a random fluctuation at a phase transition, one expects that its
statistical average should vanish, and thus it is natural to assume
$K(\tau_0)=0$, or in any case $K(\tau_0)\ll 1$ if one understands $K$
more as a variance than as a mean value. Provided it does not grow
much during its subsequent evolution, the linear equation of state
should thus remain a good approximation.

The parameter space for which the chirality $K$ is stable/unstable can
be easily understood if we fix $L_\text{c}$, $v$ and $Y$ according to
their scaling solution values, and solve the equation for $K$. This
reads
\begin{equation}
\label{Keq}
    \dot{K} \tau = 2\bar\alpha K - 2 K_o,
\end{equation}
where
\begin{equation}
\label{Alpha}
\bar\alpha = \frac{v\sca k(v\sca)}{\zeta\sca \sqrt{1+Y\sca}} - n
\end{equation}
and
\begin{equation}
\label{Beta}
K_o =
\frac{(1-2 \rho_\textsc{a}) A Y\sca}{\zeta\sca
\sqrt{1+Y\sca }} + 
\frac{v\sca}{\zeta\sca} c 
\left[ g(Y\sca)-1 \right]
(1-2 \rho) \sqrt{1+Y\sca}.
\end{equation}
The general solution for the chirality then has the form
\begin{equation}
    \label{KSolut}
    K = \left[ K(\tau_0)-\frac{K_o}{\bar\alpha}\right] \left(
    \frac{\tau}{\tau_0}\right)^{2 \bar\alpha} +
    \frac{K_o}{\bar\alpha},
\end{equation}
in which the constant term should provide the scaling value $K\sca =
K_o/\bar\alpha$ providing $\bar\alpha<0$.  In order for the $K(\tau)$
to vanish asymptotically, i.e. for $K_o=0$, one must have both
$\rho_\textsc{a}=\rho_\text{\st{bias}}$ (no bias towards either charge
or current in the leakage mechanism) and either $g(Y\sca)=1$ (loops
and long strings have the same average charge or current) or
$\rho=\rho_\text{\st{bias}}$ (charge and current losses of equal
amounts), as expected. If either of these parameters takes a different
value, then a fixed amount of chirality will be produced as the
network evolves. This is in fact of no consequence as far as the other
variables are concerned, and should not affect the scaling solution,
unless $K_o/\bar\alpha$ is large enough that it drives the model into
its non-linear regime.

The case $\bar\alpha\to 0$ must be treated separately. It yields a
logarithmic divergence in time for $K(\tau)$ unless $K_o= 0$.

If $\bar\alpha>0$ (respectively $\bar\alpha < 0$), $K(\tau)$ grows
(resp.  decays) and the distance to the equilibrium chirality
increases (resp. decreases).  One can plug the analytic solutions
given by Eq.~\eqref{ScalinLinStd}, \eqref{SolutionNonSt} and
\eqref{SolutNonTrLeak} to \eqref{Alpha} to see that these lead to
\begin{equation}
\label{Alpha2}
\bar\alpha = \frac{2 k_o}{c_o+k_o} - n= \frac{\alpha}{1-\alpha}n,
\end{equation}
where we used Eq.~\eqref{GrowAlpha}.  The expression \eqref{Alpha2}
for the radiation epoch ($n=1$) is always positive when $k_o>c_o$.
Bearing in mind that scaling solutions with non-trivial current
amplitude are attractor for $\zeta\sca$, $v\sca$ and $Y\sca$ only when
$k_o >c_o $ (as shown in section \ref{Stability}), we can conclude
that $K$ does not have a stable scaling solution when the current
amplitude is non-trivial $Y\sca \neq 0$ for a linear charge leakage or
a linear perturbation of $g$.

By studying the behaviour of $\bar\alpha$, we conclude from
Eq.~\eqref{Alpha} that $K(\tau)$ is growing in the parameter space
where the non-trivial charge is an attractor. Hence, there is no
stable scaling solution with non-trivial charge $Y\sca$ due to the
instability of the chirality $K$.  This could be due to the behaviour
of the parameters $\rho$ and $\rho_\textsc{a}$ in
Eqs.~\eqref{EqOfMotMacroLin2}, which we have not explored so far.

As was described in Ref.~\cite{MPRS}, the parameter $\rho$ was
introduced as a possible skew factor of charge/current loss due to the
production of loops. In other words, each loop contains some amount of
charge $Q$ (timelike contribution) or current $J$ (spacelike
contribution), and this amount is controlled by the parameter
$\rho$. If it is constant, each loop contains the same relative
amounts of charge and current, so that $\rho=\frac12$ implies that
each loop contains the same amount of charge and current. If we
initially have a small deviation form chirality, i.e. $K(0) \neq 0$,
and there is a non-vanishing $Y$, the string network will go away from
chirality and $K(\infty)$ will tend to diverge from zero.

A more generic behavior should allow for the increase or decrease of
charge loss in comparison to current if either one of them dominates
over the other initially.  If $Q^2>J^2$, the string network should
more likely lose charge rather than current, and this can be modeled
through
\begin{equation}
\label{Rho}
\rho = \frac{1}{2} - s K = \frac{1}{2} - s \left( Q^2 - J^2 \right),
\end{equation}
where $s>0$ is a constant. Plugging \eqref{Rho} in the time evolution
equations for $Q$ and $J$, given by Eqs.~(42) of Ref.~\cite{MPRS}, one
sees that if $Q^2>J^2$, $\rho$ yields a positive contribution to $J^2$
and a negative one to $Q^2$, meaning that the averaged charge per
string length decreases and the current per string length increases.
Substituting $\rho$ given by \eqref{Rho} into \eqref{EqOfMotMacroLinD}
and again extracting the parameter $\bar\alpha$, still using the
analytic form \eqref{SolutionNonSt}, one obtains
\begin{equation}
\label{Alpha3}
\bar\alpha = \frac{k_o (2-n) - n c_o}{2 B^2 (c_o+k_o)c_o}
\left( B - 2s + 4B s  \right),
\end{equation}
where we have assumed that $k_o (2-n) - n c_o > 0$ since only in that
parameter region does the scaling solution with non-trivial current
amplitude exist. We also used \eqref{RestrBRelation} for $b$, assuming
$\frac14 < B < \frac12$.  To ensure that $\alpha$ is negative, one
then needs to set $s > B/[2(1-2B)]$, a relation which is always
possible, demanding $s>\frac14$, provided $B\neq 0$.

An example of a time-dependent solution of Eq.~\eqref{EqOfMotMacroLin}
with $g$ and $\rho$ given by \eqref{GLinear} and \eqref{Rho}, and
$k(v)=k_o$, is shown in figure \ref{Fig:evol}.B, for the radiation and
matter eras. We find that the solution is indeed dynamically driven to
a non-vanishing constant charge in the radiation era, charge which
subsequently vanishes during matter domination; the chirality also
vanishes $K \to 0$.  One should keep in mind however that our model
concerns average values, so that the actual superconducting cosmic
string network may contain both timelike and spacelike and/or chiral
current-carrying strings.

An analogous treatment can be done for $\rho_\textsc{a}$, i.e. one can
set
\begin{equation}
\label{RhoA}
\rho_\textsc{a} = \frac12 - s_\textsc{a} K = \frac12 -
s_\textsc{a} \left( Q^2 - J^2 \right),
\end{equation}
with a similar interpretation as for $\rho$: if the string network
contains a timelike contribution larger than the spacelike one, it is
more likely that charge, rather than current, might escape the
network. Substituting \eqref{RhoA} in \eqref{EqOfMotMacroLinD2},
setting $g=1$, and assuming scaling for $L_\mathrm{c}$, $v$ and $Y$,
one finds that
\begin{equation}
\begin{gathered}
\label{Alpha4}
\bar\alpha = \frac{ k_o (2-n) - n c_o}{(c_o+k_o) c_o}
\left( 1 - 4 s_\textsc{a} Y\sca \right),
\end{gathered}
\end{equation}
where $Y\sca$ is given by Eq.~\eqref{SolutNonTrLeak}.  One sees that
there are choices of $s_\textsc{a}$ for which one can ensure
stability, i.e. with $\bar\alpha<0$ for $K$. An example of such an
evolution for the system \eqref{EqOfMotMacroLin2} is shown on
Fig. \ref{Fig:evol}.D, again showing a charged scaling during
radiation domination followed by a transition to the NG scaling when
matter kicks in.

Let us conclude this section by stating that we have shown the CVOS
system \eqref{EqOfMotMacroLin2} to have only NG uncharged, $Y\sca =
0$, fixed points for $n \geq 2$: provided the phenomenological
parameters are in the relevant domain, the charge $Y$ is dynamically
driven to a non vanishing constant $Y\sca\not= 0$ while the radiation
sources the Universe expansion, it subsequently decays after the
radiation-to-matter transition. This is illustrated
Fig.~\ref{Fig:evol}.B and \ref{Fig:evol}.D. It is interesting to note
that analogous classes of solutions were also found in the chiral
superconducting and wiggly string cases
\cite{Oliveira:2012nj,Almeida}, so one could be led to conjecture that
it might be a generic behavior for a more universal current-carrying
string equation of state \cite{MPRS}.

\section{The standard momentum parameter}
\label{Seckv}

So far, we made the simplifying assumption that the momentum parameter
was constant, $k\to k_o$, although it is known, at least in the NG
case for which there are numerical simulations allowing the evaluation
of $k$, that it is dependent on the RMS velocity $k=k(v)$. Lacking
similar simulations for the current-carrying case, one cannot decide
whether it should or could depend on either the charge $Y$ and/or the
chirality $K$, and we will therefore stick with the simplifying
assumption that $k$ depends on neither.

At scaling, with all the relevant functions of time reaching constant
values, one can safely assume $k\to k_o$, although one might still ask
if the solution for the scaling variables is a good approximation to
the exact case. Further, there is {\sl a priori}, for an arbitrary
functional dependence $k(v)$, no particular reason why its derivative
should remain small around the scaling solution, which could
potentially undermine our previous approximate stability analysis. In
this section, we accordingly discuss the validity of the results
obtained in the previous sections in the case for which the momentum
parameter does, as in the usual NG case, depend on the RMS
velocity. As discussed above, we will assume that it retains the
standard form given by \cite{Martins:2000cs}
\begin{equation}
k_\textsc{ng} (v) =\frac{2\sqrt{2}}{\pi} \frac{1-8 v^6}{1+8 v^6}
\left( 1-v^2\right) \left( 1+2 \sqrt{2} v^3 \right)\,.
\label{MomentumFunct}
\end{equation}
It should be emphasized that this form has been obtained by comparison
with NG simulations, but we could also modify it further to a slightly
different velocity dependence, such as that which found more recently
by comparison with high-resolution field theory simulations
\cite{CorreiaMartins}. However, the specific form of the
velocity-dependent function should not impact our results too
significantly, as long as the function $k(v)$ is monotonic (which is
true for both of these forms), not least because we are seeking
scaling solutions where the velocity is a constant, i.e.\ in which $k$
itself becomes a constant. We also want to emphasize that the shape
\eqref{MomentumFunct} for $k(v)$, plotted e.g.\ as Fig.~3 of
Ref.~\cite{MPRS}, is varying only very slowly over a wide range of
velocities, and therefore one might expect the stability analysis
assuming $k(v) \sim k_o$ to be mostly valid over this regime of
velocities.

For $k(v)$ given by Eq.~\eqref{MomentumFunct}, the CVOS model needs to
be studied numerically. However, we expect the most important
difference with the cases discussed in sections
\ref{Stability}-\ref{StabilityK} to stem from the fact that the
constant ratio $k_o/c_o$, that mattered so much for the stability
analysis, is now promoted to a dynamical variable. As a result, for a
given set of underlying parameters, we will find that is now possible
for two different attractors to simultaneously exist. In particular,
the solution $Y\sca=0$, which is always present, could then be always
stable, thereby restricting the possible set of initial conditions
leading to a charged network: if the mechanism by which the current
originally forms is such that at that time, the system is close to the
$Y=0$ solution, it would then return to that state, and even though in
principle a charged scaling solution exists, it may never be actually
reached. What we shall find is that, for a given set of
phenomenological parameters allowing for a charged scaling
configuration, the space of initial conditions for $v$, $\zeta$ and
$Y$ contains two clearly separated regions, shown e.g. in
Fig.~\ref{Figure:PhaseDiagrLin2}, in which the subsequent evolution is
fully deterministic, leading inevitably to a charged or an uncharged
network.

\begin{figure}[h!]
\begin{center}
\includegraphics[scale=0.4]{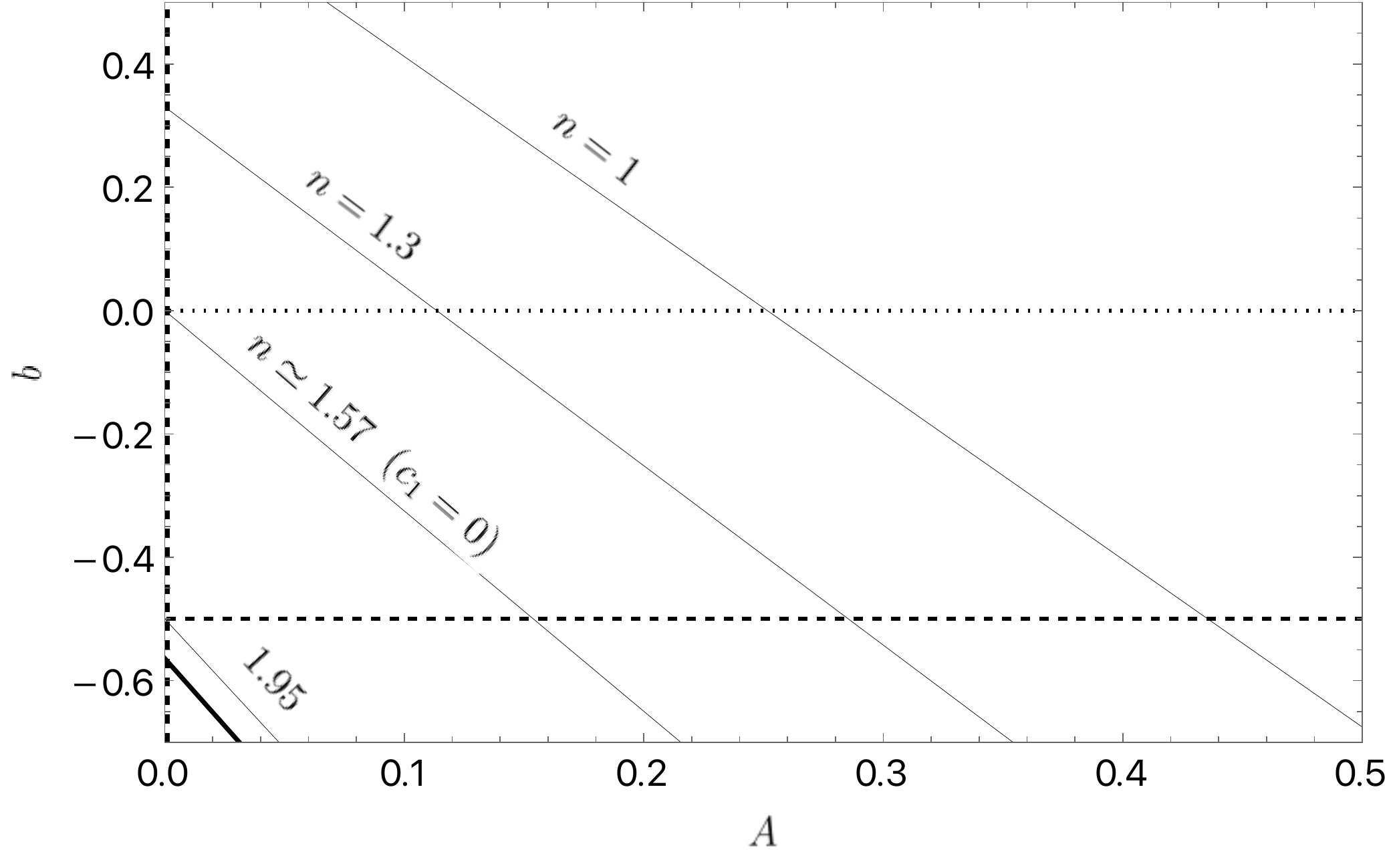}
\caption{\label{Figure:bA} Boundary lines, in the $(A,b)$ plane,
  between regions of potentially charged and certainly uncharged
  network solutions, for various expansion rates $n$, assuming a
  velocity-dependent momentum parameter $k(v)$ given by
  Eq.~\eqref{MomentumFunct}. On the top right side of each line, only
  solutions having $Y\sca=0$ are attractors, whereas on the bottom
  left side, attractor solutions with $Y\sca=0$ as well as
  $Y\sca\neq0$ exist, that may be reached depending on the initial
  conditions (see Fig.~\ref{Figure:PhaseDiagrLin2}); the dashed
  vertical axis $A=0$ is excluded from the region for which there is a
  non trivial solution. The best fit
  (\ref{BestFitNum1}--\ref{BestFitNum2}), in the displayed case for
  which $c_o=0.23$, has $\alpha_1=1.616$, $\beta_1 = -0.932$,
  $\gamma_1=1.226$, $\alpha_2 = -2.624$, $\beta_2 = -0.095$ and
  $\gamma_2 =4.198$. The thick line in the bottom-left corner
  represents the matter-dominated era $n=2$, and the thin line
  labelled ``$1.95$'' is that for which the $Y\sca\neq0$ attractor
  demands $b<-\frac12$ (dashed line), in contradiction with the
  constraint \eqref{b12}.  }
\end{center}
\end{figure}

We repeat the analysis of Sec.~\ref{Stability}, by numerically
searching for all possible fixed points with physically meaningful
values, e.g.\ satisfying $0<v\sca<1$, $\zeta\sca>0$ and $Y\sca\geq 0$.
As a first result in the varying momentum parameter case, we recover
the conclusion that a charge loss mechanism, such as the leakage in
Eq.~\eqref{ChargeLeakage}, is again needed to modulate growing charge
solutions to ensure there is a stable non-trivial equilibrium point
with $Y\sca \neq 0$.

\begin{figure}
\begin{center}
\includegraphics[scale=0.46]{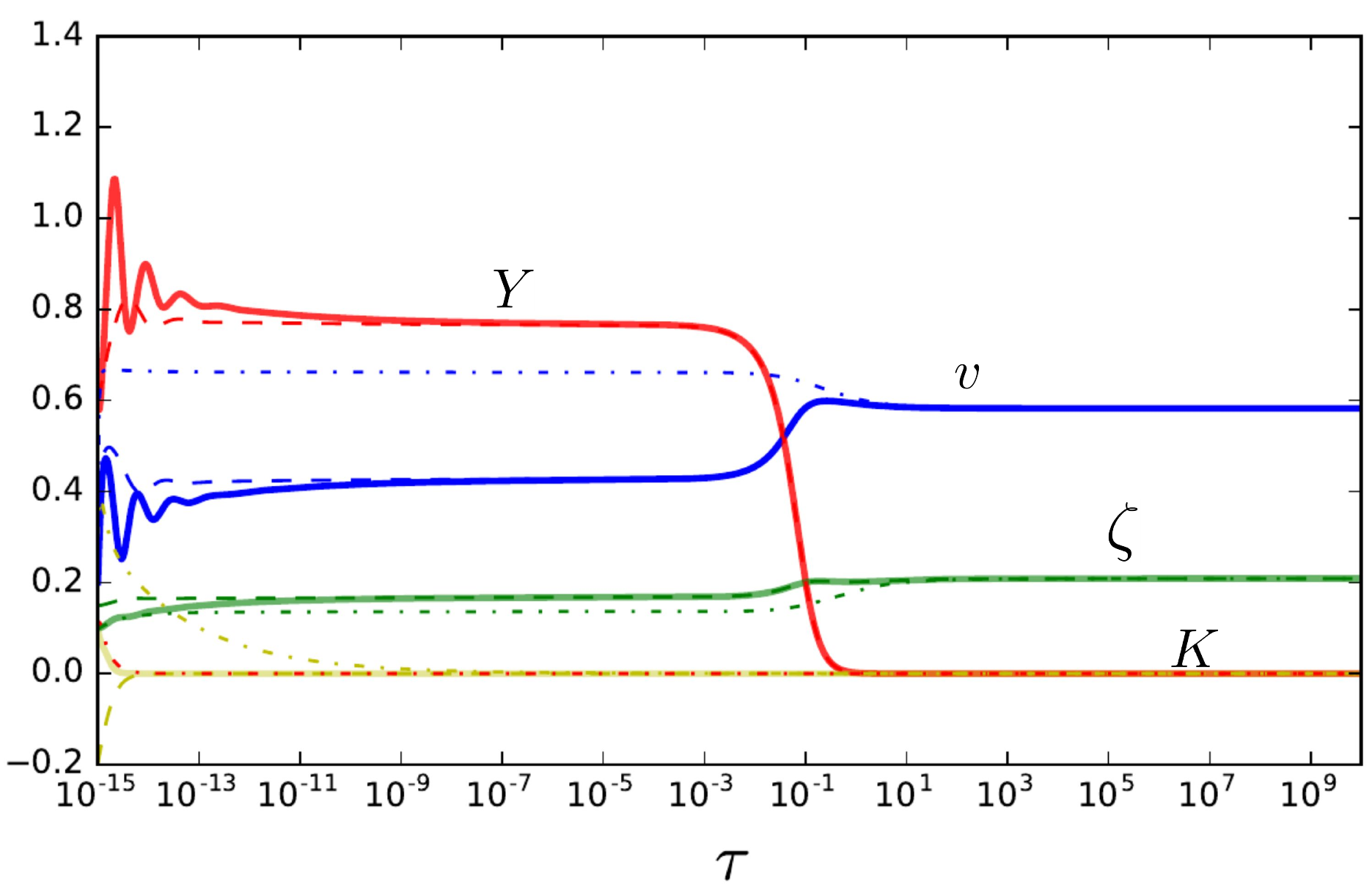}
\caption{\label{Figure:Mat-Rad_kv2} Time evolution of the velocity
  $v$, charge $Y$ and $\zeta$ for parameter values $c_o=0.23$, $b=0$,
  $A=0.25$, $\beta=0$, $s=1$, and $s_\textsc{a}=1$, during the
  radiation and matter epochs, for different initial conditions: the
  solid and the dashed trajectories correspond, in
  Fig.~\ref{Figure:PhaseDiagrLin2}, to initial conditions inside the
  dark (blue) region leading to a non-vanishing scaling charge, and
  the dash-dotted line to the NG network configuration.  Even in the
  cases for which the $Y\sca\not=0$ attractor is reached while $n=1$
  before the radiation-to-matter transition, the subsequent dynamics
  then drives the solution to an uncharged on ($Y\sca=0$ attractor) in
  the matter era.}
\end{center}
\end{figure}

\begin{figure}[h!]
\begin{center}
\includegraphics[scale=0.58]{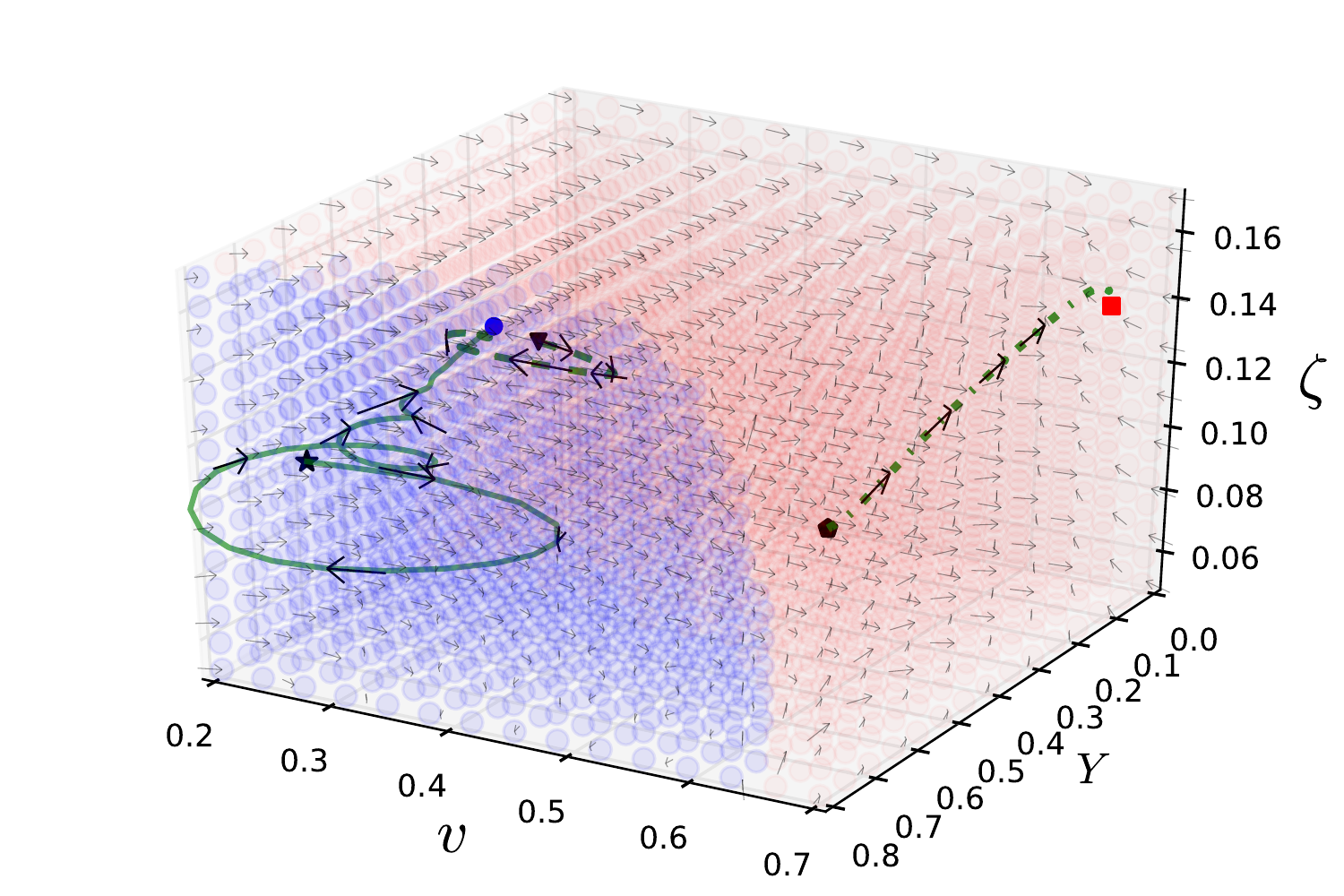}
\caption{\label{Figure:PhaseDiagrLin2} Phase diagrams showing the
  nature of the fixed points, i.e. leading to a charged $Y\sca\neq0$
  or NG $Y\sca=0$ scaling configuration in the radiation era
  ($n=1$). Setting initial conditions in the blue (dark) region leads
  to a trajectory ending into a charged scaling solution, whereas
  starting in the red region leads inevitably to a NG network.  The 3
  curves shown represent 3 different trajectories, those also shown in
  Fig.~\ref{Figure:Mat-Rad_kv2} and for the same parameter set.  The
  scaling value with non-trivial charge leads to the momentum
  parameter $k(v\sca) \approx 0.81$, which lies inside the area with
  $Y\sca \neq 0$ of Fig.~\ref{Figure:Leakage}, while that with
  vanishing charge yields $k(v\sca) \approx 0.18$, lying inside the
  area with $Y\sca = 0$ of Fig.~\ref{Figure:Leakage}.  The black dots
  show the initial conditions, chosen close to the boundary surface
  and on both sides.  }
\end{center}
\end{figure}

Having found that modifying the chopping parameter $c$ by means of the
charge dependence \eqref{ChopEff} does not lead (on its own) to new
scaling solutions for the constant momentum parameter $k_o$ case, we
assume that this is still the case when it takes the form of
Eq.~\eqref{MomentumFunct}: were a different scaling solution to exist,
upon reaching it, the parameters would be mostly constant, and our
previous analysis, showing no $Y\sca\neq0$ solution to exist, should
apply. For this reason, we assume initially that $\beta=0$ in what
follows and explore the 2-dimensional parameter space $(A,b)$.

In comparing with our previous analysis with $k(v) = k_o$, the
important difference with the dynamical momentum parameter $k(v)$ is
that for the same fixed parameters it is possible to have two scaling
attractors: Nambu-Goto and charged. The configuration that is actually
dynamically reached is determined by the network's initial conditions.

Fig.~\ref{Figure:bA} shows an example of the region in the $(A,b)$
plane where a $Y\sca\neq0$ attractor solution exists with the chopping
efficiency set to $c_o=0.23$. As expected from the previous
discussion, the regions with one solution $Y\sca=0$ only and those
with two solutions, including both $Y\sca=0$ and $Y\sca\neq0$
(depending on the initial conditions), are neatly separated.
Repeating the analysis for various values of the expansion rate $n$ as
shown in Fig.~\ref{Figure:bA}, we affirm the conclusion that the
matter-dominated era cannot sustain a charged network
configuration. As $n$ is lowered, the region in parameter space which
can sustain a charged solution increases: numerical investigations
reveal that the boundary curves between the charged and uncharged
networks are well approximated by the straight lines
\begin{equation}
b = c_1(n) + c_2(n),
\label{BestFitNum1}
\end{equation}
where
\begin{equation}
c_i(n) = \alpha_i(c_o)+ \beta_i(c_o) n^{\gamma_i(c_o)},
\label{BestFitNum2}
\end{equation}
so that $c_1(n_0)=0$ yields the value of $n=n_0$ such that only for $b
< 0$ do non-trivial solutions exist (for $n>n_0$, see
Sect.~\ref{gbias}). The constraint \eqref{b12} also implies that
$\exists n_\text{crit}$ such that $c_1(n_\text{crit}) = -\frac12$, and
$\forall n \geq n_\text{crit}$, $Y\sca=0$ whatever the initial
conditions.

These dynamics which change in different cosmological eras and depend
on initial conditions can give rise to an interesting phenomenology,
especially as a charged network crosses the radiation-matter
transition.  For a set of values of $A$ and $b$ located between the
boundary lines $n=1$ and $n=n_\text{crit}$ (as illustrated in
Fig.~\ref{Figure:bA}), one may find initial conditions for $v$,
$\zeta$ and $Y$ such that the dynamics leads to the $Y\sca\neq0$
attractor during the radiation-dominated era, followed, after the
radiation-to-matter transition, by a new time evolution towards the
$Y\sca=0$ attractor. An example of such an evolution during the
radiation-to-matter transition, with initial conditions appropriately
chosen, is illustrated in Fig.~\ref{Figure:Mat-Rad_kv2}. The
corresponding configuration space trajectories for the radiation epoch
are shown in Fig.~\ref{Figure:PhaseDiagrLin2}, emphasizing how initial
conditions define the choice of scaling behavior.  The region of
initial conditions that leads to the scaling solution with non-trivial
current amplitude depends on the parameters $A$ and $b$ (and possibly
on $\beta$ if included, although the phenomenology will be quite
similar).

\begin{figure}
\begin{center}
\includegraphics[scale=0.58]{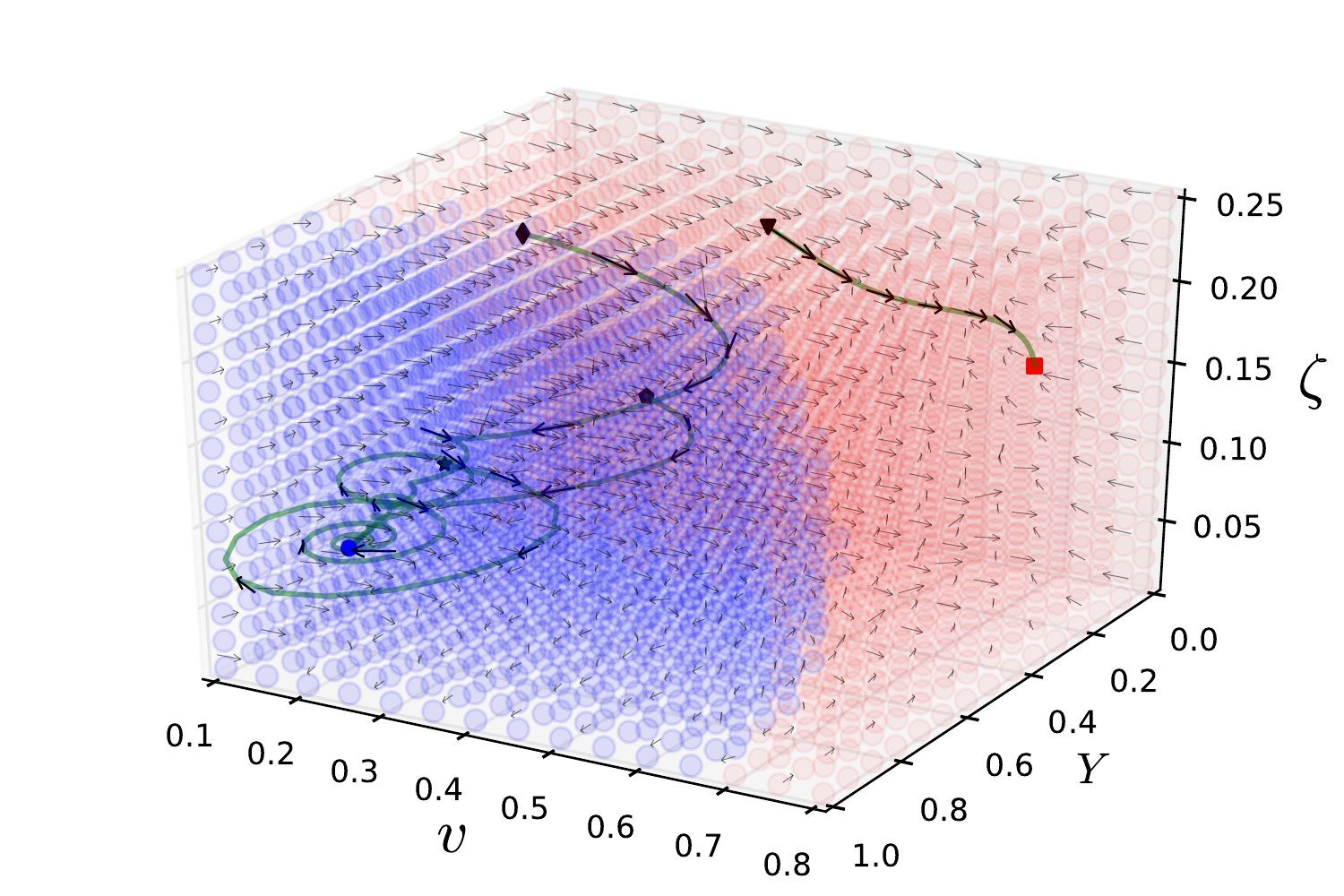}
\includegraphics[scale=0.58]{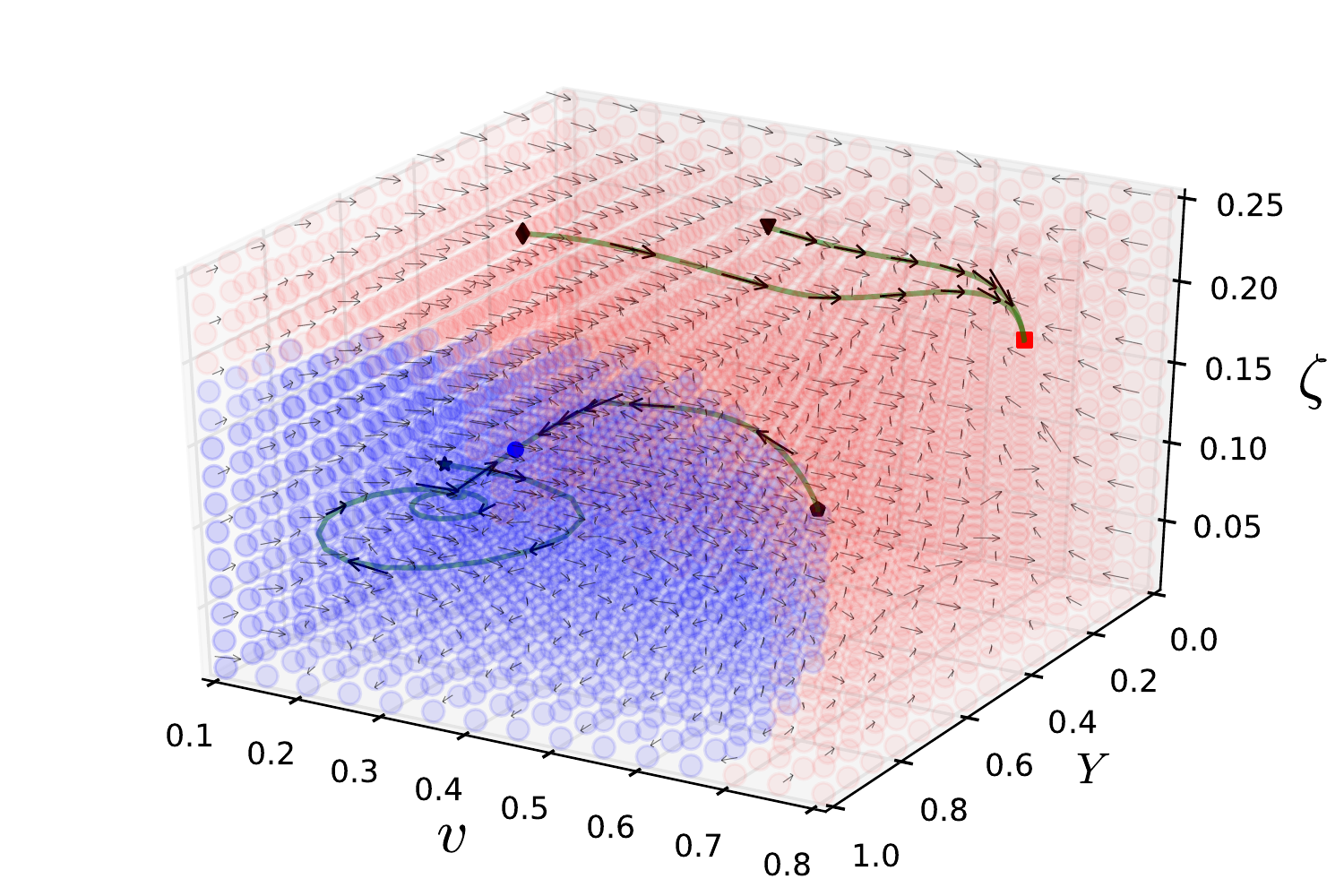}
\includegraphics[scale=0.58]{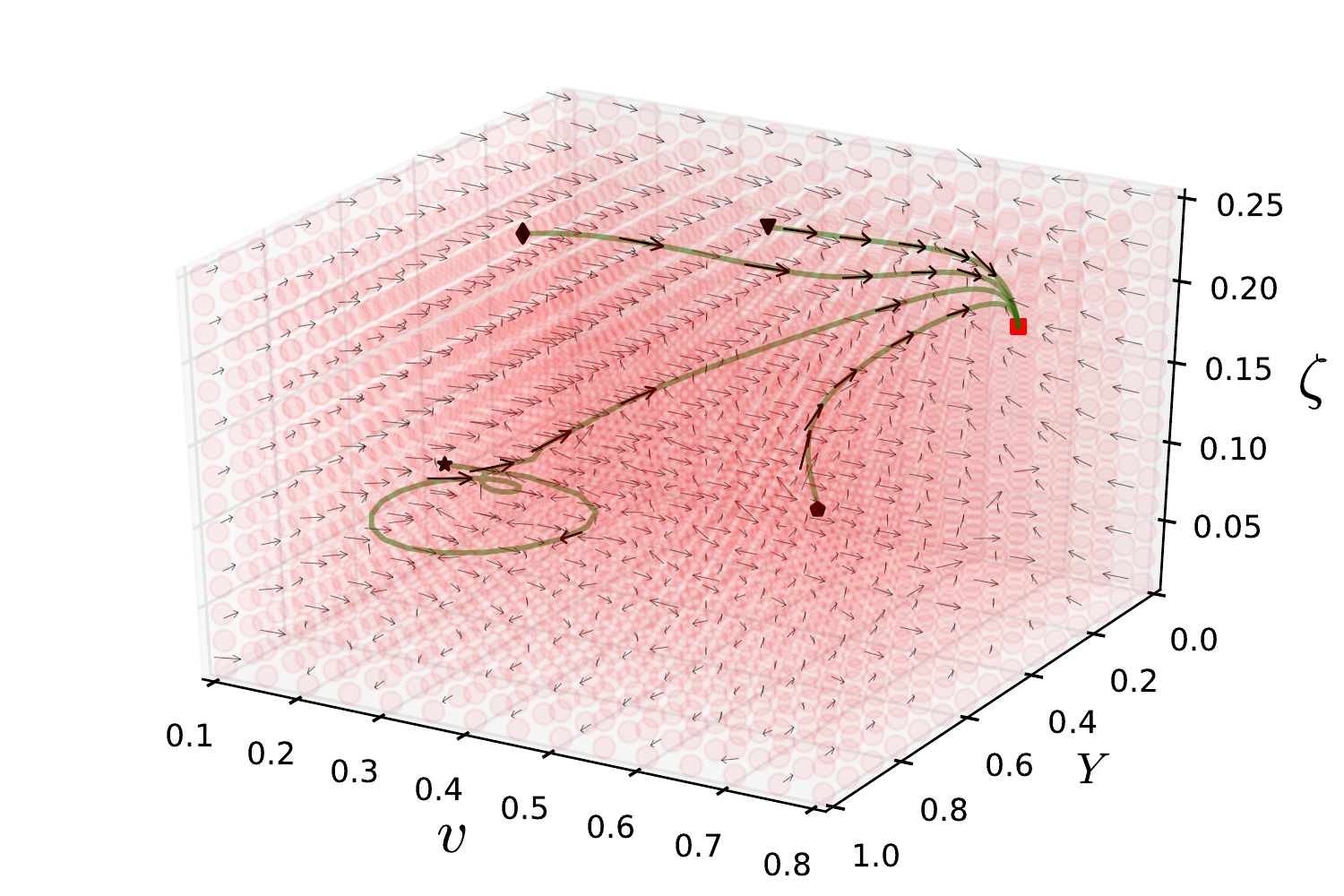}
\caption{\label{Figure:PhaseDiagrLin3} Same as
  Fig.~\ref{Figure:PhaseDiagrLin2} for $n=1$ (top panel), $n=1.2$ (mid
  panel) and $n=1.3$ (bottom panel) with identical parameters except
  $A\to 0.15$. As expected from the discussion above and shown in
  Fig.~\ref{Figure:bA}, the available region (in blue/dark) for which
  initial conditions for $v$, $\zeta$ and $Y$ may lead to a non
  trivial charge $Y\sca\neq0$ scaling solution is getting smaller as
  $n$ increases.  The charged configuration attractor points are shown
  as dark blue circles, and the standard uncharged configuration as a
  red square.  The initial conditions shown as \emph{triangle},
  \emph{pentagon}, \emph{diamond} and \emph{star} are the same for all
  3 graphs. In this example, for $n\geq 1.3$, there are no charged
  solution, and the would-be attractor becomes a repeller.  }
\end{center}
\end{figure}

Fig.~\ref{Figure:PhaseDiagrLin3} illustrates the behaviour of the
regions in configuration space having both charged and uncharged or
only uncharged attractors. As discussed above, for smaller $n$, the
larger the region in which $Y\sca\neq 0$ can be reached.  Above a
threshold value ($n=1.3$ in this case), there is no longer any charged
configuration available. This explains why the typical time evolution
of a superconducting string network will be to first form a charged
but chiral state, rapidly evolving, after the radiation-matter
transition, into a non current-carrying standard network.

\section{Conclusions}

The time evolution of of a superconducting cosmic string network in
the framework of the extended VOS model involves, on top of the usual
RMS string network velocity $v$ and characteristic length
$L_\textsc{c} = \zeta \tau$, not only the chirality parameter
$K=Q^2-J^2$ (the difference between the integrated squared charge $Q$
and current $J$), but also the overall charge amplitude $Y=\frac12
(Q^2+J^2)$.  By examining the time evolution, we found that the linear
equation of state approximation, expected to be valid for small
currents, arguably leads to a natural evolution towards the chiral
case; this is for arbitrary initial conditions and in the absence of
some bias between the charge and current loss mechanisms. In other
words, starting with a non-trivial value for $K$ describing the
distance to chirality (with $K>0$ spacelike current and $K<0$
timelike), we found that in all regimes studied, i.e., both in the
radiation and the matter dominated epochs, this chirality parameter
tends to rapidly vanish, leading to an effective chiral model.  In any
case, the essential decoupling of the chirality $K$ from the other
parameters $v, L_\textsc{c}$ and $Y$ meant that their dynamics could
be considered separately.

The CVOS dynamical system is found here to possess two distinct
scaling solutions, if the expansion rate is sufficiently slow, that
is, typically including the cosmologically relevant radiation
dominated epoch.  The specific scaling solution that emerges depends
on the initial conditions of the network: the standard NG scaling
solution with no currents or charges arises from small initial
currents, while there is a possible non-trivial attractor with
non-vanishing current amplitude $Y$, which emerges from relatively
large initial currents (and/or low string velocities).  We leave for
further investigation the question of whether such high-current
initial conditions are feasible in realistic scenarios.  On the other
hand, for fast expansion rates, including the matter dominated epoch,
we find there is no non-trivial scaling solution and all initial
conditions are unavoidably driven towards the NG uncharged scaling
solution.

The non-trivial attractor with non-vanishing charge $Y$ is
characterised, as compared to the NG solution, by a lower RMS velocity
and a smaller characteristic length, leading to an overall larger
energy of the network (as compared to uncharged strings), but with
part of the energy being in the charge/current contribution.  In order
for this scaling solution to be stabilized, however, one needs to
introduce a charge leakage mechanism so that some charge can move away
from the network, e.g. due to local losses at high curvature string
regions.  Under these circumstances, the predicted scaling charge $Y$
will be a strongly model-dependent quantity, depending on
microphysical dynamics. On the other hand, in the absence of such
charge leakage, the attractor solution would be one with a
continuously growing charge/current in which the network eventually
becomes `frozen' with $v=0$. We note that such non-scaling growing
solutions have also been found for chiral superconducting strings
\cite{Oliveira:2012nj} and, in a slightly different context, for
wiggly strings \cite{Almeida}.  However, this growing charge solution
is unlikely to be physically realised because, as discussed, we expect
charge loss mechanisms to intervene to yield a scaling solution.

The fact that we obtain either charged or Nambu-Goto attractor
solutions raises an issue that deserves further consideration,
namely the question of initial conditions for the charge $Y$ and
chirality $K$. As we have shown, the space of initial condition
shows two separated regions, at least for slowly expanding
universes, in which the dynamics leads to very different outcomes:
either a charged scaling solution or one without any charge. Unlike
the Nambu-Goto case, current-carrying strings do exhibit some
sensitivity to their initial conditions. Nevertheless, for fast
expansion, we recover the usual (charge-free) scaling solution so
initial conditions again become irrelevant.

From the point of view of early universe cosmology, our main result is
that charges and currents could play a significant role in the
evolution of string networks in the radiation era (quite probably
leaving behind astrophysical fingerprints) but, after the
radiation-to-matter transition, the time evolution drives the charge
amplitude $Y$ towards zero, that is, the uncharged NG scaling
configuration. In particular, this could lead to a radically different
spectrum of gravitational waves in both eras, e.g.  potentially
suppressing short wavelength signals. For this reason, one could
anticipate that realistic string networks, possessing internal degrees
of freedom, might generate enhanced observational signatures that are
linked to the radiation-matter transition.

Finally, we note that we have obtained these results in the framework
of the small current limit for which the linear equation of state
should be a valid approximation. While we expect this to be
representative of string forming phase transitions in the early
universe, a question that may legitimately be asked is whether our
result of no charge or current in the matter era is a consequence of
the smallness of the current in the radiation era, and of the specific
use of the linear equation of state, or a fully generic result that
will also apply for other, possibly more realistic (Witten-like)
worldsheet actions? The fact that growing charge solutions clearly
exist for slow expansion rates, and previous knowledge of the
evolution of chiral superconducting and wiggly string networks
\cite{Oliveira:2012nj,Almeida} leads us to suspect that this behaviour
is qualitatively generic: regardless of the equation of state, a fast
enough expansion rate will always suffice to make charges and currents
disappear. Quantitatively, the question is then whether the frontier
between slow and fast expansion rates depends on the equation of state
(or only on the available energy loss mechanisms), and whether the
matter dominated epoch is always on the fast (no current) side of this
frontier. A more detailed study of this issue is left for future work.

\acknowledgments

This work was financed by FEDER---Fundo Europeu de Desenvolvimento
Regional funds through the COMPETE 2020---Operational Programme for
Competitiveness and Internationalisation (POCI), and by Portuguese
funds through FCT - Funda\c c\~ao para a Ci\^encia e a Tecnologia in
the framework of the projects POCI-01-0145-FEDER-031938,
POCI-01-0145-FEDER-028987 and PTDC/FIS-PAR/31938/2017,
PTDC/FIS-AST/28987/2017.

PP was hosted at Churchill College, Cambridge, partially supported by
a fellowship funded by the Higher Education, Research and Innovation
Dpt of the French Embassy to the United-Kingdom.

PS acknowledges funding from the STFC Consolidated Grants ST/P000673/1
and ST/T00049X/1.

IR also wants to express his gratitude to Juliane F. Oliveira for
useful discussions on the stability of scaling solutions.

\bibliography{references}

\end{document}